\documentclass{emulateapj}

\newcommand{\mincir}{\raise-2.truept\hbox{\rlap{\hbox{$\sim$}}\raise5.truept \hbox{$<$}\ }}
\newcommand{\magcir}{\raise-2.truept\hbox{\rlap{\hbox{$\sim$}}\raise5.truept\hbox{$>$}\ }}
\def\lsim{~\rlap{$<$}{\lower 1.0ex\hbox{$\sim$}}}
\def\bsim{~\rlap{$>$}{\lower 1.0ex\hbox{$\sim$}}}
\newcommand{\beq}{\begin{equation}}
\newcommand{\eeq}{\end{equation}}
\newcommand{\bdm}{\begin{displaymath}}
\newcommand{\edm}{\end{displaymath}}
\newcommand{\bea}{\begin{eqnarray}}
\newcommand{\eea}{\end{eqnarray}}
\newcommand{\bt}{\begin{tabular}}
\newcommand{\et}{\end{tabular}}

\usepackage{graphicx}
\usepackage{times}
\usepackage{epsfig} 

\def\etal{{\it et al.}}

\def\etal{{\rm\,et \ al. }}
\def\hmpc{{\rm Mpc} \ h^{-1}}
\def\hkpc{{\rm kpc} \ h^{-1}}
\def\kms{{\rm km} \ {\rm s}^{-1}}

\shorttitle{Studying the WHIM with GRBs}
\shortauthors{Branchini et al.}

\begin{document}

\title{Studying the WHIM with Gamma Ray Bursts}

\author{E.~Branchini\altaffilmark{1}, 
E.~Ursino\altaffilmark{1},   
A. ~Corsi\altaffilmark{2}, 
D.~Martizzi\altaffilmark{1}, 
L.~Amati\altaffilmark{3}, 
J.W. den Herder \altaffilmark{4},  
M.~Galeazzi \altaffilmark{5},  
B.~Gendre\altaffilmark{6},  
J.~Kaastra\altaffilmark{4,14}, 
L.~Moscardini\altaffilmark{7,15},  
F.~Nicastro\altaffilmark{8}, 
T.~Ohashi\altaffilmark{9},  
F.~Paerels\altaffilmark{4,10},  
L.~Piro\altaffilmark{2}, 
M.~Roncarelli\altaffilmark{11,8}, 
Y.~Takei\altaffilmark{4,16} 
and M.~Viel\altaffilmark{12,13}
}  
\email{branchin@fis.uniroma3.it}
\altaffiltext{1}{Dipartimento di Fisica, Universit\`a degli Studi ``Roma Tre'' via della Vasca Navale 84, I-00146 Roma, Italy}
\altaffiltext{2}{INAF-Istituto di Astrofisica Spaziale Fisica Cosmica, Via del Fosso del Cavaliere 100, I-00133 Roma, Italy}
\altaffiltext{3}{INAF-Istituto di Astrofisica Spaziale e Fisica Cosmica Bologna, via P. Gobetti 101, I-40129
Bologna, Italy}  
\altaffiltext{4}{SRON Netherlands Institute for Space Research, Sorbonnelaan 2, 3584 CA Utrecht 
The Netherlands}  
\altaffiltext{5}{Physics Department of University of Miami,
319 Knight Physics Building, Coral Gables, FL 33164, U.S.}
\altaffiltext{6}{Laboratoire d'Astrophysique de Marseille/CNRS/Universit\'e de Provence, 38 Rue Joliot-Curie, 
13388 Marseille CEDEX 13, France}
\altaffiltext{7}{Dipartimento di Astronomia, Universit\`a degli Studi di Bologna, via Ranzani 1, I-40127 Bologna, Italy}
\altaffiltext{8}{INAF-Osservatorio Astronomico di Roma,via Frascati 33, I00040 Monteporzio-Catone (RM), Italy}
\altaffiltext{9}{Department of Physics, School of Science, Tokyo Metropolitan University, 1-1 Minami-Osawa, Hachioji, Tokyo 192-0397, Japan}
\altaffiltext{10}{ Columbia Astrophysics Laboratory and Department of Astronomy, Columbia University, 550 West 120th Street, New York, NY 10027, U.S.}   
\altaffiltext{11}{ Centre d'\'Etude Spatiale des Rayonnements (CESR) - CNRS, Observatoire Midi-Pyr\'en\'ees, 9 avenue du Colonel Roche, 31028 Toulouse Cedex 04, France}
\altaffiltext{12}{ INAF-Osservatorio Astronomico di Trieste,  via Tiepolo 11, I-34131 Trieste, Italy}
\altaffiltext{13}{INFN/National Institute for Nuclear Physics, Via Valerio 2, I-34127 Trieste, Italy}
\altaffiltext{14}{Astronomical Institute, University of Utrecht, Postbus 8000, 3508, TA Utrecht, The Nederlands}
\altaffiltext{15}{ INFN/National Institute for Nuclear Physics, Sezione di Bologna, Via Berti Pichat 6/2, I-40127, Bologna, Italy}
\altaffiltext{16}{Institute of Space and Astronautical Science,
Japan Aerospace Exploration Agency,
3-1-1 Yoshiodai, Sagamihara, Kanagawa 229-8510, Japan}

\newpage

\begin{abstract}
We assess the possibility to detect and characterize the physical state of the
missing baryons at low redshift by analyzing the X-ray absorption spectra of
the Gamma Ray Burst [GRB]  afterglows, measured by a
micro calorimeters-based detector  with 3
eV resolution and 1000 cm$^2$ effective area and capable of fast repointing,
similar to that on board of the recently proposed X-ray
satellites {\small {\small EDGE}} and {\small {\small  XENIA}}.
For this purpose we have analyzed mock absorption spectra extracted from
different hydrodynamical simulations used to model the properties of the
Warm Hot Intergalactic Medium [WHIM].
These models predict the correct abundance of OVI absorption lines
observed  in UV and satisfy current X-ray constraints.
According to these models space missions like
{\small EDGE} and {\small  XENIA} should be able to detect $\sim$ 60
WHIM absorbers per year through the OVII line.
About 45 \% of these have at least two more detectable lines in addition to OVII
that can be used to
determine the density and the temperature of the gas.
Systematic errors in the estimates of the gas density and temperature
can be corrected for in a  robust, largely model-independent fashion.
The analysis of the GRB absorption spectra collected in three years
would also allow to measure the cosmic mass density of the WHIM with $\sim$ 15  \% accuracy,
although this estimate  depends on the WHIM model.
Our results suggest that GRBs represent a  valid, if not preferable, alternative to Active Galactic Nuclei
to study the WHIM in absorption. The analysis of the absorption spectra nicely complements
the study of the WHIM in emission  that  the spectrometer proposed for
 {\small EDGE} and {\small  XENIA}  would be able to
 carry out thanks to its high sensitivity and large field of view.
\end{abstract}


\section{Introduction}

According to the standard cosmological scenario, baryonic matter provides a small fraction, about 4-5 \%, of the total cosmic energy content. While baryons are to first order dynamically negligible in shaping the large-scale structure of the Universe, they have been the essential component in creating stars, galaxies, planets and the life in the Universe and constitute the only component that interacts with electro-magnetic radiation. As such, they allow to study in great detail, through diverse
observations, the physical conditions for the formation and evolution of cosmic structures.

As a matter of fact  the baryon density in the Ly-$\alpha$ forest at
redshift 2-3 (Weinberg \etal  1997, Rauch 1998) 
accounts for a large fraction of the baryon mass as inferred from
the predictions of the  Big Bang Nucleosynthesis combined with
the measured abundances of the light elements (Burles and Tytler 1997, 1998)
and, independently, with the temperature fluctuations of the cosmic microwave background (Spergel \etal 2003,
Komatsu \etal 2008).
On the other hand, an observational census of cosmic baryons in the local Universe
(Fukugita, Hogan \& Peebles 1998,  Fukugita \& Peebles 2004, Danforth \& Shull 2005)
shows that  at $z \sim 0$ only about 50 \% of the baryons have already been detected.
The content of stars in galaxies and the interstellar medium provides only about 8 \% of the overall baryonic budget; the hot plasma with temperature of at least $10^7 \ {\rm K}$ contained in the potential wells of galaxy clusters adds an additional 11 \%. Finally, the diffuse gas with temperature below $10^7 \ {\rm K}$, that constitutes the local Ly-$\alpha$ forest,  provides a further 29 \% (Bregman 2007).

Hydrodynamical numerical simulations have provided
a  solution to this paradox within the framework of the 'concordance' $\Lambda$CDM model.
They predict that about half of the total baryonic content of the Universe at
$z<1$ should be in a  warm-hot phase with temperature in the range
$ 10^5$ - $10^ 7\  {\rm K}$, distributed in a web of tenuous filaments 
(Cen \& Ostriker 1999a).
As such, the WHIM represents the dominant
component of baryons in the low redshift  Universe.
Its thermal evolution is primarily driven by shock heating from gravitational perturbations breaking onto  unvirialized cosmic structures, such as large scale filaments and virialized structures, such as galaxies, groups and clusters of galaxies. Feedback effects related to the  nuclear activity or to
galactic superwinds from galaxy and star formation also play a role, especially in high density environments (e.g. Cen \& Ostriker 2006). These baryons are so hot and so highly ionized that they can only emit or absorb in the far UV and soft However, being the typical density of the WHIM lower than that of the hot Intra-Cluster Medium it is extremely hard to detect. 
Yet, it must contribute to the soft X-Ray Background. The  importance of this contribution clearly depends on the physical conditions of the WHIM (e.g. Voit \& Bryan 2001,  Bryan \& Voit 2001). The exquisite spatial resolution of the {\small Chandra} X-ray observatory has now allowed to resolve a large part, more than 80 \%, of the cosmic X-ray background  at soft energies into the contribution of point-like sources (mostly Active Galactic Nuclei - AGN hereafter). Hickox \& Markevitch (2007a, 2007b) recently analysed the Chandra Deep Fields North and South. After removing the contribution from X-ray, optical and IR sources and carefully characterizing the instrumental background and non-cosmological foreground, they computed the residual background. According to their analysis, this residual background
in the 0.65-1 keV band cannot be accounted for by extrapolating the source number counts to lower fluxes.
Instead, it comes interestingly close to the  theoretical predictions of WHIM emission
 based on cosmological hydrodynamical simulations that include realistic description of star formation and SN feedback (e.g. Roncarelli \etal  2006). If due to  soft X-ray emission from the diffuse gas, this residual
 X-ray background would represent the first, though indirect, detection of the missing baryons in emission.

Clearly,  the study of the unresolved soft X-ray background just provides an upper limit to the amount of diffuse gas and indirect indications of its physical properties.
The best chance to detect the WHIM is by observing the emission or absorption lines of highly ionized  element like C, N, O, Ne, and possibly Mg and Fe.
The  strongest lines expected correspond to the outer K (X-ray band) and L (Far UV band) shell transitions from H-like, He-like and Li-like  Oxygen ions: the
OVIII 1s-2p  X-ray doublet [from now on simply OVIII], the OVII  1s-2p X-ray resonance line
[OVII, hereafter] and the OVI UV doublet [OVI-UV].
These measurements, however, are at the limit of current instrumentation's capability.

So far, the best WHIM detection has been provided by the analysis of the UV spectra of 31 AGN by Danforth \& Shull (2005). They detected  
40 OVI-UV absorption systems counterparts to  Ly-$\alpha$  absorbers, whose 
cumulative number per unit redshift
above a given equivalent width agrees with the latest theoretical predictions (Cen \& Fang 2007).
Most  of the Ly-$\alpha$ lines associated to these absorbers, however, are too  narrow to be produced by  gas at temperatures larger than  $ 10^5 \  {\rm K}$.
This means  that either these lines are associated to the local Ly-$\alpha$ forest, and thus their contribution to the total baryon budgets has been already accounted for (Tripp \etal 2008), or that two different gas phases are found at the same location: a photo-ionized gas responsible for the Ly-$\alpha$ absorption and
shock-heated WHIM responsible for the  OVI lines.
Assuming that all OVI-UV lines are indeed produced by a multi-phase
gas, one obtains an upper limit to the mass density associated to these
absorbers which corresponds to 7-10 \% of the missing baryonic mass
(Danforth \& Shull  2005 and  Tripp \etal  2006).
Alternatively, one could search for broad  Ly-$\alpha$ lines in the UV spectra, i.e. lines with a width parameter  $b \ge 40 \  \kms $ associated with gas  hotter than $ 10^5 \  {\rm K}$.
The recent searches by  Richter \etal (2006) and Stocke, Danforth \& Shull (2008)
showed that the few broad absorbers detected so far contribute to 2-10 \% of the mass of the missing baryons.

All these studies show that while some of the missing baryons have been revealed through OVI lines,
the vast majority of them is  too hot to produce significant OVI opacity. Therefore  the bulk of the
WHIM can only be revealed and studied in the soft X-ray band.
In emission, the diffuse signal detected in correspondence with overdensity of galaxies
has been interpreted as the signature of the WHIM by Zappacosta \etal (2005) and by Mannucci \etal (2007).
In absorption, among the several claimed WHIM detections  at $z>0$ (e.g. Fang \etal 2002, Mathur \etal 2003), the most convincing one has been provided by Nicastro \etal (2005) who detected several WHIM absorption lines in the X-ray spectrum of the extraordinarily bright Blazar Mrk 421.
Yet, the statistical significance of these detections has been questioned by Kaastra \etal (2006) through Monte Carlo simulations. Moreover,
the absorption lines originally seen in the spectrum obtained with the Low Energy Transmission Grating Spectrometer on board of {\small Chandra} were not detected in a spectrum
of the same object subsequently 
obtained through a longer exposure with the Reflection Grating Spectrometer on {\small XMM-Newton} (Rasmussen \etal 2006).
The very existence of this unsettled controversy demonstrates the importance of performing a dedicated experiment to  detect the missing baryons in the X-ray band.
Even more, going beyond the mere detection of the WHIM, the characterization of  its thermal and chemical properties,  will open an unexplored territory for the study of the interaction and co-evoultion of the diffuse phase of cosmic baryons and the  stellar population.

The possibility of detecting the WHIM in the absorption spectra of X-ray bright objects
like AGN has been thoroughly discussed  from a theoretical
perspective in a series of  studies. While Chen \etal (2000) and Cen \& Fang (2006), to mention a few examples, addressed the problem in a generic cosmological setup, other authors
(e.g.  Kravtsov \etal 2002, Klypin \etal 2003, Viel \etal 2005) focused on the 
the gas distribution in the local Universe. In addition, other works have specialized
the treatment to some specific satellites,
like  {\small XMM-Newton} (Kawahara \etal 2006) or the
proposed {\small DIOS} (Yoshikawa \etal 2003, Yoshikawa \etal 2004),
{\small XEUS} and {\small Constellation-X} (Viel \etal 2003) satellites.
A comprehensive review of the current and future instrumentation for the study of the WHIM
can be found in Paerels \etal (2008).
In this work we focus on the possibility of revealing the missing baryons signature in the X-ray spectra
of the GRB afterglows.
The idea of using GRBs as bright X-ray beacons to study the
physical properties of the intervening gas and trace its evolution out to very large redshifts has
been first proposed by Fiore \etal (2000). Here we push this idea further and exploit the
recent SWIFT results on the rate of bright GRBs
to estimate the probability to detect the WHIM and to characterize its physical status
in the context of the recently proposed {\small EDGE} and {\small  XENIA}
space missions.
In addition, rather than relying on some specific theoretical  prediction,
we consider different WHIM models obtained from hydrodynamical simulations
to account for theoretical uncertainties.

The characteristics of the detectors on board of {\small EDGE} and {\small  XENIA}
can be found in Piro \etal (2008), in Kouvelioutou \etal (2008) and
 at the web address http://projects.iasf-roma.inaf.it/EDGE.
Gamma-Ray Bursts are localized by a Wide Field Monitor, a hard
X-ray detector covering about 1/4 of the sky. The position derived
on board is used to command  a satellite autonomous slew, that
enables the acquisition of  the GRB location by the Wide Field
Spectrometer, an X-ray telescope equipped with high spectral
resolution TES microcalorimeters,  in 60 s.

Here we list for reference the  capabilities of the two
instruments that are more relevant to this work:

\begin{itemize}

\item Wide Field Spectrometer [WFS]: Energy range $0.2-2.2$ keV.
Energy resolution $\Delta E =3$ eV (goal: 1 eV) at 0.5 keV
(TES microcalorimeters). Field of view  ${\rm FOV} \ 0.7^{\circ} \times 0.7^{\circ} $.
Effective area A=1000 cm$^2$  at 0.5 keV.  Angular resolution: $\Delta \theta =2$ arcmin.

\item Wide Field Monitor [WFM]: Energy range $6-200$ keV. Energy
resolution $\Delta E/E =3\%$ at 100 keV. Field of view  ${\rm FOV} \ >2.5$ sr. 
Effective area A=1000 cm$^2$  at 100 keV. Location accuracy: $\Delta \theta =3$ arcmin.
\end{itemize}

This work is organized as follows. In Section 2 we  
describe our WHIM models and 
discuss
the characteristics of the
hydrodynamical simulations that we have used to extract the
mock X-ray absorption spectra.
In Section 3 we analyze the statistical properties of the
absorption lines in the mock X-ray spectra, characterize the
physical properties of the absorbing  material and investigate
the statistical correlations among the absorption lines.
In Section 4 we quantify the probability of detecting WHIM lines
in the  GRB afterglow spectra and estimate the number of
detections expected with a satellite like {\small  XENIA} or {\small EDGE}.
In Section 5 we investigate the possibility of studying the physical status
of the WHIM and to measure its cosmological abundance.
We discuss our main results and conclude in Section 6.

\section{Modeling the WHIM}
\label{sec:whimodel}

The  best WHIM models currently available are based on numerical hydrodynamical simulations
and  provide robust predictions for the  cosmological WHIM abundance
as a function of redshift,  the filamentary appearance of its
large-scale distribution and  its thermal state
(Cen \& Ostriker 1999a, 1999b, Chen \etal 2001, Cen \& Ostriker 2006 to provide a few examples).
The latter is mainly determined by hydrodynamical shocks resulting from the build-up of cosmic structures at scales that have  become nonlinear. Additional, non-gravitational heating mechanisms that depend on the little known feedback processes following star formation in galaxies.
However, they are relevant only
 in high density environments in which the role of        radiative cooling must  also be taken into account.
In this work we do not rely on a single WHIM model that includes
sophisticated  recipes to treat feedback effects.
Instead, in an attempt to bracket model uncertainties,
we use two different hydrodynamical
simulations to develop three WHIM models with different phenomenological
treatments of the non-gravitational heating mechanisms,
star formation and feedback effects.

\begin{figure}

\centering

\epsfig{file=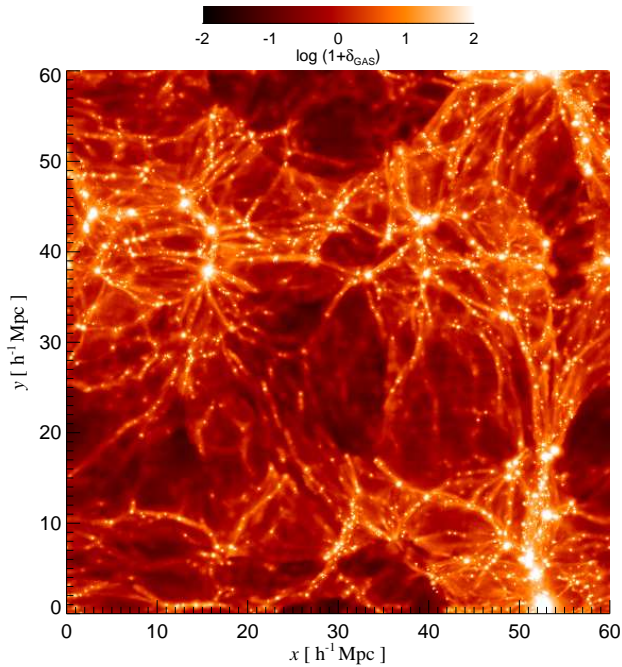,width=0.5\textwidth}

\caption{The spatial distribution of the gas overdensity $\delta_{\rm g}$
 in a slice of
thickness 6 $\hmpc$ (comoving) at $z=0$
extracted from the simulations described in Pierleoni \etal (2008)
and centered
around the largest cluster in the simulation box (to be found in bottom--right part of the panel)}

\label{fig:sim1}

\end{figure}

The first WHIM model (dubbed model $M$ in Table~1)
relies on the same simulations described in
Pierleoni \etal (2008).
This numerical experiment, of which we only consider the outputs at $z\le1$,
was performed with  the  GADGET-2 Lagrangian code (Springel 2005)
in a computational box of size $60h^{-1}$ comoving Mpc, loaded with
$480^3$ Dark Matter [DM] and $480^3$ gas particles.
The Plummer equivalent  gravitational softening was set
equal to 2.5 ${\rm kpc} \ h^{-1}$ in comoving units for all particles.
This choice of parameters represents
a good compromise between box size and resolution and allows to investigate 
the WHIM  properties down to the Jeans scale. 
Radiative cooling and heating processes were
followed according to Katz, Weinberg \& Hernquist (1996).
The simulation assumes
the same X-ray and UV background at $z=0$ that  Viel \etal (2003) used
to determine the ionization state of the intergalactic gas.
The X-ray background is not computed self consistently but is modeled 
{\it a posteriori} as
$I_{X}=I_{X}^{0}(E/E_{X})^{-1.29}\exp{(-E/E_{X}) }$, with
$I_{X}^{0}=1.75 \times 10^{-26}$ erg cm$^{-2}$ Hz$^{-1}$ sr$^{-1}$
s$^{-1}$ and $E_{X}=40$ keV,  in agreement with observations
(Boldt 1987; Fabian \& Barcons 1992; Hellsten \etal 1998).
The UV background is of the form $I_{UV} = I_{UV}^0 (E/13.6\rm{eV})^{-1.8}$, with
$I_{UV}^{0}=2.3 \times 10^{-23}$ erg cm$^{-2}$ Hz$^{-1}$ sr$^{-1}$
s$^{-1}$ , which agrees within a factor of 2 with the estimate  of Shull \etal (1999)
and shows no appreciable differences with the intrinsic Haardt-Madau like spectrum
(Haardt \& Madau 1996) used in the simulation run.
The cosmological parameters of this simulation are similar
to the {\it WMAP}-1 year values (Spergel \etal ~2003)
and refer to a flat $\Lambda$CDM  'concordance'  model with  $\Omega_{\Lambda}=0.3$,
$\Omega_b=0.0463$, $H_0=72 \  \kms \ {\rm Mpc}^{-1}$, $n=0.95$ and $\sigma_8=0.85$.
In Fig.~\ref{fig:sim1} we show the gas distribution at $z=0$
of a slice of thickness 6~$\hmpc$ (comoving)  extracted from the computational box.

The star formation criterion simply
converts all gas particles whose temperature falls below $10^5$ K and
whose density contrast is larger than 1000 into (collisionless) star particles.
No energy feedback from supernova explosions, metal ejection and diffusion was considered.
After determining the thermal state of  each gas particle we assign the metal
abundance in the post-processing phase according to a phenomenological metallicity-density
relation, calibrated on hydrodynamical simulations in which metal enrichment is treated self-consistently.
In model $M$  we have assumed  the
deterministic relation ${\rm Z}=\min(0.2,0.02(1+\delta_{\rm g})^{1/3})$ of Aguirre \etal (2001),
where $ {\rm Z}$  is  the gas metallicity in 
solar units and
 $\delta_{\rm g} \equiv \rho_{\rm g}/\langle \rho_{\rm g} \rangle-1$  is the gas 
overdensity with respect to the mean,  $\langle \rho_{\rm g} \rangle$.
To model the WHIM properties we  have used  all available outputs of the hydro simulation
with $z\le1$ using the stacking technique described in Roncarelli \etal (2006).

The second simulation that we use has been carried out by Borgani et al. (2004)
using the same GADGET-2 code. The box size is larger
that in the previous case and consists of a comoving cube of  192 $\hmpc$ side
 containing $480^{3}$+$480^{3}$
dark matter and gas particles. The Plummer-equivalent gravitational softening
is $\epsilon$=7.5 $h^{-1}$ kpc at $z=0$, fixed in physical units between $z=2$ and $z=0$ and
fixed in comoving units at earlier times. The cosmological model  is a flat  $\Lambda$CDM with $\Omega_{\Lambda}=0.7$,  $\Omega_{\rm b}=0.04$,  $H_0=70  \ \kms \ {\rm Mpc}^{-1}$ and  $\sigma_8=0.8$.
Differently from  the previous case, this simulation
includes a treatment of the main  non-gravitational physical processes that influence the physics of
the gas: {\it i)} the star formation mechanism, that is treated  adopting a
sub-resolution multiphase model for the interstellar medium
(Springel \& Hernquist 2003), {\it ii)} the feedback from SN\ae\ including the effect of weak
galactic outflows and
 {\it iii)} the radiative cooling assuming zero metallicity and heating/cooling from the uniform, time-dependent, photo-ionising UV background by Haardt \& Madau (1996).

We have used all available  outputs of the simulation with $z\le1$ to construct
two more WHIM models that differ by
their metallicity-density relation.
The rational behind specifying the metal abundance in the post-processing phase
rather than adopting the one computed self consistently in the simulation is that
the simple metal diffusion mechanism in the SPH code 
reproduces  the correct metallicity in  high density regions  like the intracluster medium, 
but underestimates the metal abundance in low density environments like the WHIM.
The first model obtained from the simulation of Borgani \etal (2004), 
dubbed $B1$ in Table~1, assumes the deterministic metallicity-density relation
of Croft \etal (2001):  ${\rm Z}=\min \ (0.3,0.005(1+\delta_{\rm g})^{1/2})$.
Several hydrodynamical experiments (e.g. Cen and Ostriker 1999a, 2006), have revealed that
the relation between  ${\rm Z}$ and  $\delta_{\rm g}$
is characterized by a very large, non-Gaussian  scatter around the mean,
reflecting  the fact 
that the thermal state and the metal content of the gas  are not uniquely determined by the local density
but, rather, by its dynamical, chemical and thermal history.
In model $B2$ we account for this stochasticity
by enforcing in our hydrodynamical simulation
 the same scatter found by
Cen \& Ostriker (1999a,1999b) in their numerical experiment.
To do this we have measured the density-metallicity  relation
in the simulation outputs of  Cen \& Ostriker (1999a) 
and constructed a 2D probability distribution in the  ${\rm Z}-\delta_{\rm g}$ 
plane by counting particles in each bin, at various redshifts.
The resulting 2D grid   provides a probability distribution that we implement in a Monte Carlo 
procedure  to assign  metallicity to the  gas particles 
in the Borgani \etal (2004) simulation
with known density.

The rationale behind considering three different WHIM models is to 
 provide some indication of the theoretical uncertainties that
reflect  our little understanding of the effects of stellar feedback from star- and galaxy formation, chief among which is of course the metal enrichment. 
The latter is so important in determining the observational properties of the WHIM that model
predictions can be obtained by specifying gas metallicity in the
post-processing phase come close to those computed self-consistently
in the simulations (Chen \etal 2001).
In what follows we mainly focus on model $B2$ that we regard as more realistic. However, we also  consider the two models $B1$ and $M$, which adopt ideal metallicity models directly 
used in previous analyses  that have appeared in the literature.
We stress that these models do not bracket all unknows in the theoretical calculation
since they adopt simple recipes for metal enrichment and stellar  feedback processes and
ignore departures from  equilibrium that, for example, have been considered in the 
more sophisticated  WHIM model of Cen \& Fang (2006).

\begin{deluxetable}{ccccccl}
\tablecaption{WHIM Models}
\label{tab1}
\tablehead{
\colhead{Model} &
\colhead{L$_{\rm Box} $} &
\colhead{${\rm N}_{\rm DM}+{\rm N}_{\rm gas}$} & 
\colhead{$\epsilon$} &
\colhead{$\Omega_b$} &
\colhead{$\sigma_8$} &
\colhead{Z-$\rho_{\rm g}$ relation}
} 
\startdata
$M$ & 60 &  $400^3+400^3$& 2.5 &  0.046 & 0.85 & Z $\propto \rho_{\rm g}^{1/3}$ \\  
$B1$ & 192 &   $480^3+480^3$ & 7.5  & 0.04 & 0.8 & Z $\propto \rho_{\rm g}^{1/2}$ \\ 
$B2$ & 192 &  $480^3+480^3$ & 7.5 & 0.04 & 0.8 & Z-$\rho_{\rm g}$ + scatter \\  
\enddata
\tablenotetext{}{Column 1: Model name. Column 2: Box size (in comoving $\hmpc$).
Column 3: Number of DM and gas particles. 
Column 4: Plummer equivalent gravitational softening (in $\hkpc$).
Column 5: Baryon density parameter.
Column 6: Mass variance at 8 $\hmpc$. Column 7: Metallicity-Density  relation adopted.
}
\end{deluxetable}

After having simulated the metals and their distribution in the box and after having specified
the metagalactic   UV and X-ray background, we have used the photoionization code
CLOUDY (Ferland {\it et al.} 1998) to compute the ionization states of 
these metals under the hypothesis of hybrid collisional and photoionization equilibrium.
The resulting ionization fractions  turns out to be very similar to that obtained by Chen \etal (2003) and shown in their Figs. 2 and 3 that implies that our OVII fraction in  regions of moderate overdensity
is different from that computed by Kawahara \etal  (2006) using a modified SPEX code (see their Fig. 1). 
The density of each ion was obtained through: $n_I(x)= n_H(x) X_I Y_{Z_{\odot}} {\rm Z}$, where $X_I$ is the ion fraction determined by CLOUDY that depends on gas temperature, gas
density and ionizing background, $n_H$ is the density of hydrogen atoms, Z is the metallicity of the element in solar units, and $Y_{{\rm Z}_{\odot}}$ is the solar abundance of the element (Asplund \etal 2005). 

The ions responsible for the most prominent X-ray absorption lines are listed in Table 2together with the energy of the transition and the oscillator strength. These are the X-ray lines that we simulate in our mock absorption spectra. In addition, to compare model predictions with existing data we also consider the OVI UV doublet (at 11.95 eV, 12.01 eV), which  has been observed in the absorption spectra taken by  {\small FUSE} and {\small HST} (Tripp \etal 2000, Savage \etal 2002,  Danforth \& Shull 2005).

To check the validity of the of  hybrid ionization-collisional equilibrium hypothesis, 
we have  computed the typical recombination time in the WHIM $t_{rec}=(n_{WHIM}\, \alpha_{rec})^{-1}$, with $\alpha_{rec}$ the recombination rate. For the two more relevant ions OVII and OVIII, in the temperature ranges considered here we find $\alpha_{rec}\sim 2\times 10^{-11}$ cm$^3$ s$^{-1}$ (Mazzotta \etal 1998), where both the dielectronic and radiative recombinations are taken into account. This means that the
Hubble time is larger than the recombination time  for a density $\ge10^{-7}$ cm$^{-3}$. Thus, for a large fraction of the WHIM, the approximation of photoionization equilibrium should be reasonable.
Indeed, as it has been shown by Yoshikawa and Sasaki (2006), departures from equilibrium do not dramatically alter the equivalent widths [$EW$ hereafter] of the OVIII and OVII  absorption lines.

\begin{deluxetable}{lcc}
\tablecaption{Simulated WHIM Ions.}
\label{tab2}
\tablehead{
\colhead{Ion} & \colhead{Energy [eV]}  & \colhead{$f$}
}
\startdata
OVI & 11.95+12.01 & 0.19  \\ 
OVI 1s-2p & 563 & 0.53 \\  
OVII 1s-2p & 574 & 0.70 \\  
OVII 1s-3p & 666 & 0.15 \\  
OVIII 1s-2p & 654 & 0.42 \\ 
CV 1s-2p & 308 & 0.65 \\  
CVI 1s-2p  &367 & 0.42 \\  
NeIX 1s-2p &922 & 0.72 \\  
FeXVII 2p-3d &826 & 3.00 \\  
MgXI 1s-2p &1352 & 0.74\\  
\tablenotetext{}{Column 1: Ion. Column 2: Wavelength [eV]. Column 3: Oscillator Strength.}
\enddata
\end{deluxetable}

\subsection{Mock absorption spectra}

\label{sec:mockspectra}

Having determined the density of a given ion along the line of sight [LOS], the optical depth 
in redshift-space at velocity $u$ (in $\kms $) is

\begin{equation}
\tau_I(u)= {\sigma_{0,I}c\over H(z)} \int_{-\infty}^{\infty} dy \, n_{\rm I}(y) {\cal V}
\left[u-y-v_{\rm ||}^{\rm I}(y),b(y)\right]
\label{tau}
\end{equation}
where $\sigma_{0,I}$ is the cross-section for the resonant absorption which depends on the wavelength, $\lambda_I$ and the oscillator strength $f_I$ of the transition , $y$ is the real-space coordinate
(in km s$^{-1}$), ${\cal V}$ is the standard Voigt profile normalized in real-space, $b=(2k_BT/m_I c^2)^{1/2}$ is the thermal width. Ignoring the effect of  peculiar velocities, i.e. setting $v_{\rm ||}^I=0$, the 
real space coordinate $y$ and the redshift $z$ are related through $d\lambda/\lambda=dy/c$, where $\lambda=\lambda_I(1+z)$.
Since WHIM absorbers have low column densities, the Voigt profile is well approximated by a Gaussian: ${\cal V}= (\sqrt{\pi}b)^{-1}\exp[-(u-y)^2/b^2]$.  
The transmitted flux is simply ${\cal F}=\exp(-\tau)$.

\begin{figure}
\centering
\epsfig{file=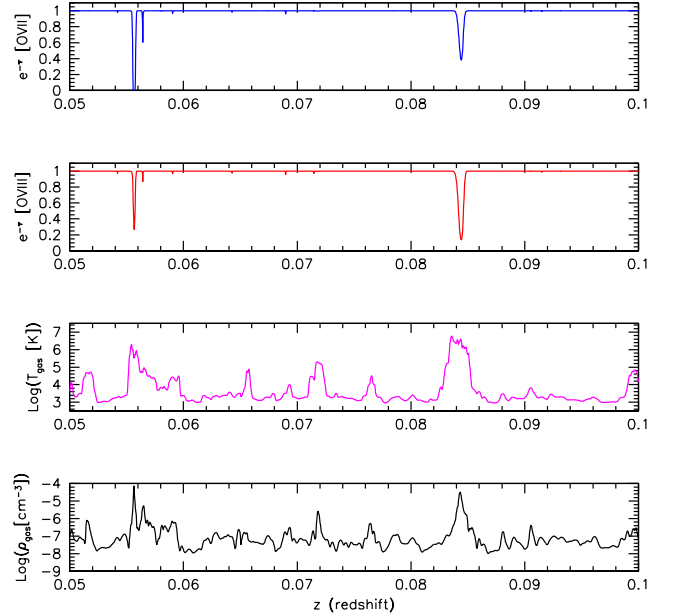,width=0.5\textwidth}
\caption{Simulated spectrum from model $B1$ for a generic LOS with two strong absorbers.
{\it From Top to Bottom}: Transmitted flux for OVII, OVIII, temperature and density along the
LOS as a function of  redshift.}
\label{fig:sp1}
\end{figure}

For each WHIM model we have extracted 15 independent spectra 
from the light-cones obtained by stacking all available simulation outputs. 
Mock spectra from model $M$ have a spectral resolution of $3 \ \kms$
and extend out to $z=0.5$. Spectra from models $B1$ and $B2$ have a resolution of $\sim 9 \ \kms$
and extend out to $z=1.0$. Figure \ref{fig:sp1} shows, for  model $B1$. 
and  from top to bottom, the transmitted flux in OVII, OVIII, the density and the temperature along a generic LOS, out to $z=0.1$.
Strong absorption features are found in correspondence of
density and temperature peaks. OVIII lines sample regions of higher density and temperature
than OVII lines. The typical number of absorbers  is small and they can be easily
identified in the mock spectra using a simple threshold algorithm.  The positions of the absorption
lines are identified with the minima of  the transmitted flux and their $EW$ are measured by integrating ${\cal F}(u)$  over the redshift interval $[\underline{z},\overline{z}]$ centered around the minimum, in
which  ${\cal F} \le  {\cal F}_{thr}$.
The few overlapping lines are separated in correspondence of the local maximum
between the two minima.
The threshold to identify absorption lines is ${\cal F}_{thr}=0.95$. We have checked
that results do not change significantly if one lowers the threshold to 75 \% of the 
transmitted flux.

\section{Line statistics}
\label{sec:linestatistics}

\begin{figure}
\centering
\epsfig{file=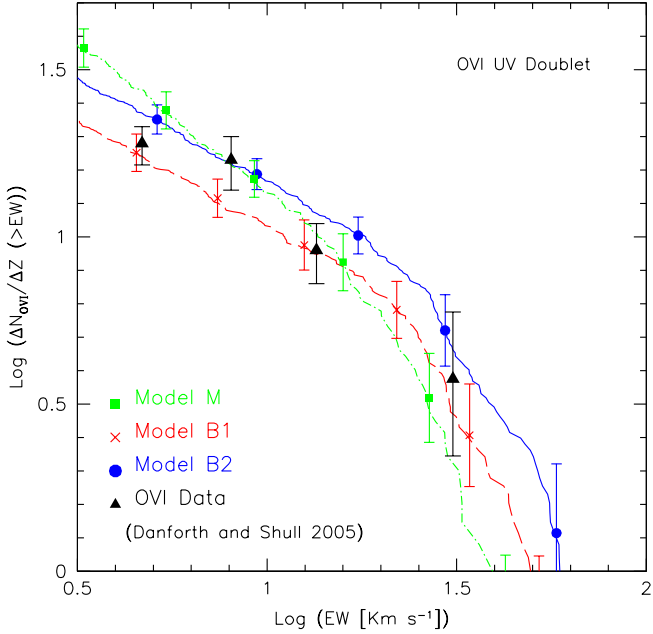,width=0.5\textwidth}
\caption{Cumulative distribution function of OVI UV doublet equivalent widths. Model
predictions vs. observations. Filled triangles with error bars: Danforth \& Shull (2005)
measurements. Filled squares and dot-dashed  green line: model $M$.
Crosses and dashed red line: Model $B1$. Filled dots and solid blue line: Model $B2$.
For each model the error bars represent the scatter in the 15 independent mock spectra. }
\label{fig:cum}
\end{figure}

To check the reliability of our WHIM models we first compare our theoretical predictions with
the only WHIM unambiguous detections available so far.
In Figure~\ref{fig:cum} the different symbols connected by continuous lines show
the cumulative distributions of  OVI UV doublet $EW_{\rm s}$ predicted by our models.
Error bars represent the scatter among the different  mock spectra.
Filled triangles with error bars show the cumulative  $EW$ distribution of the OVI absorbers
measured by Danforth \& Shull (2005). Over most of the  $EW$
range, all of the considered WHIM models are consistent with data
within the errors.
The spread in model predictions reflects  theoretical uncertainties.
The larger number of OVI lines in model $B2$ results from
the scatter in the density-metallicity relation that systematically increases the probability
of an OVI absorption line.
Model $M$ predicts a cumulative distribution steeper than models
$B1$ and $B2$. At large  $EW_{\rm s}$ the discrepancy reflects the different amount of
large scale power in the computational boxes of the parent simulations.
At small $EW_{\rm s}$, instead, the  difference originates from
the different numerical resolution of the two simulations.
In this work we focus on absorption lines with $EW$ $\ge  10 \  \kms $.
Hereafter we use different units for the line $EW_{\rm s}$. Useful
conversions between units are $EW=6.67(EW/100 \  \kms)(\lambda/20 \  {\rm \AA}) \
{\rm m\AA} =0.33(EW/100 \ \kms)(E/1 \ {\rm keV}) \ {\rm eV}$.

\begin{figure}
\centering
\epsfig{file=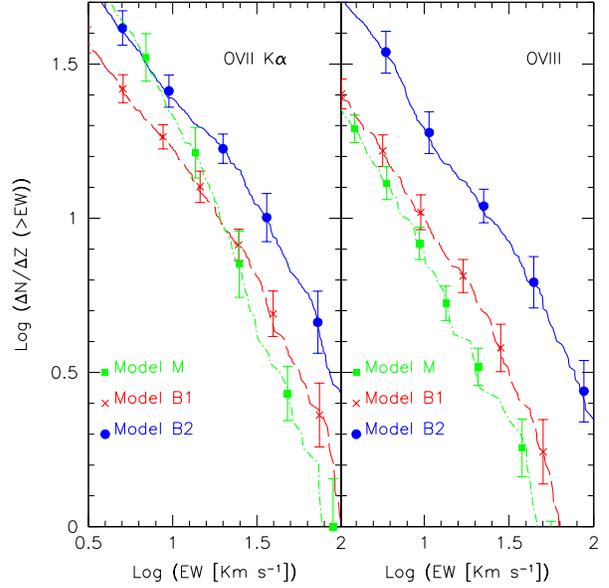,width=0.47\textwidth}
\caption{Cumulative distribution function of absorption lines $EW_{\rm s}$.
Model predictions.
Left: cumulative distribution function of OVII.
Right: cumulative distribution function of OVIII.
Filled squares and dot-dashed green line: model $M$.
Crosses and dashed red line: Model $B1$. Filled dots and solid
black line: Model $B2$.
Error bars in the model represent the scatter among
the 15 independent mock spectra. }
\label{fig:ion1}
\end{figure}

The scatter in the model predictions for the OVI line distribution function
is quite small and one may wonder whether this is also the case for the X-ray
absorption lines.  Let us consider the two strongest WHIM  lines:
OVII and OVIII. These ions probe regions of higher density
and temperature than OVI. In such environments the predicted ion abundance
is more sensitive to the different metallicity-density prescriptions adopted in the models
and thus we expect a larger spread among model predictions.
Figure~\ref{fig:ion1} shows that this is
indeed the case:  the differences among the predicted cumulative distributions
for OVII (left panel) and OVIII (right panel) are larger than in the  OVI case.
Predictions of model $B2$, that account for the scatter in the
metallictity, are consistent with those of the Chen \etal (2003)
model, in which the metallicity is drawn randomly from a lognormal
distribution with
$\langle Z/Z_{\odot} \rangle =-1.0,
 \ \ \sigma_{log Z}=0.4$.
Model $B2$ predictions are also consistent with
those of Cen \& Fang (2006), that include the feedback effect of
galactic super-winds and account for departures from local
thermodynamic equilibrium.
We note that all our predictions, including those from models $M$ and $B1$,
 satisfy the current upper limits on the OVII and OVIII line observations.

\begin{figure}
\centering
\epsfig{file=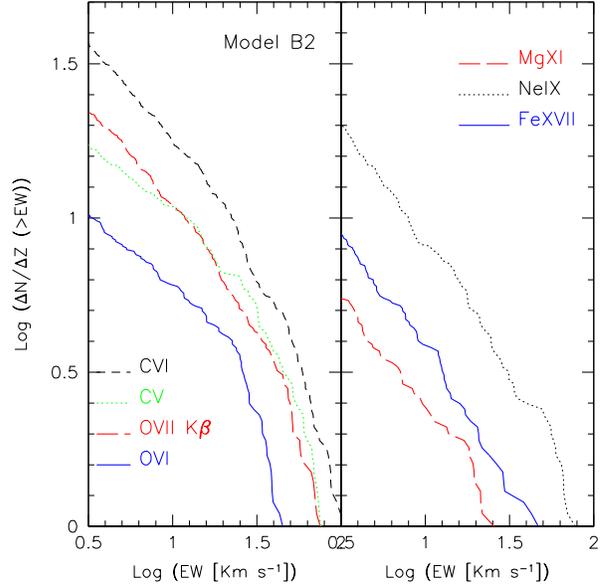,width=0.47 \textwidth}
\caption{Cumulative distribution function of absorption lines equivalent widths.
Predictions of Model $B2$.
Left: cumulative distribution function of OVI  X-ray (blue continuous) OVII 1s-3p
(red long-dashed), CV (green dotted) and CVI (black short-dashed).
Right: cumulative distribution function of FeXVII (blue solid), MgXI
(red long-dashed), and NeIX (black dotted).}
\label{fig:1ion2}
\end{figure}

In Figure~\ref{fig:1ion2} we show, for model $B2$ only, the predicted cumulative distribution
functions for the other ions listed in Table 2. As already pointed out by Chen \etal
(2003) CVI and CV are the strongest absorption lines after oxygen. However, their
transition energies of  367 and 308 eV put them in a range of poor instrument sensitivity
(see Fig.~\ref{fig:convolved}) and restrict their observability to low redshift sources.
NeIX produces the next strongest absorption together with the OVII 1s-3p line. The latter,
however, is observed in association with the stronger  OVII 1s-2p line and,
as we will discuss in the next Section,  improves the
statistical significance of the WHIM detection, but offers no further information on
the absorber's physical state.
On the contrary, the OVI line in the  X-ray band could, together with
OVII and OVIII,  be used to determine unambiguously the physics of the gas.
Unfortunately the X-ray  OVI line is typically $\sim 4$ times weaker than 
the OVII line and will be difficult to detect unless instrumental sensitivity will increase
dramatically. For this ion UV spectroscopy appears as the only 
effective option currently  available.

\subsection{Physical properties of mock absorbers}
\label{sec:linephysics}

\begin{figure}
\centering
\epsfig{file=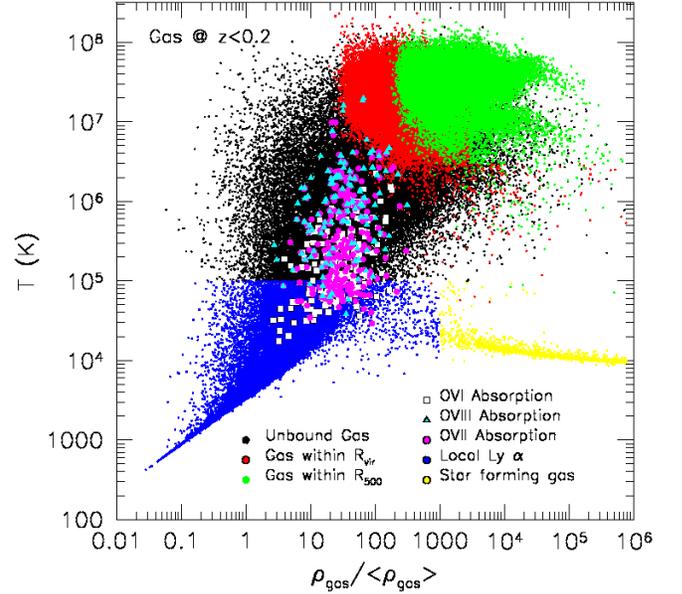,width=0.5\textwidth}
\caption{Phase space diagram of the cosmic baryons at $z < 0.2$ from the hydrodynamical simulation of Borgani \etal (2004).
Blue dots with $T\le 10^5$ and $\rho/\langle \rho \rangle \le 10^3$: gas
particles detected in the local Ly $\alpha$ forest.
Yellow dots with $T_{\rm g}\le 10^5$ K
and $\rho_{\rm g}/\langle \rho_{\rm g} \rangle \ge 10^3$  gas particles associated with
 star forming regions. Green dots with $T_{\rm g} \gtrsim 10^{6.5}$ and $\rho_{\rm g}/\langle \rho_{\rm g} \rangle \gtrsim 10^{2.5}$ gas particles in  the central part of virialized structures that are potentially observable with current X-ray satellites. Red dots  with $T_{\rm g} \lesssim 10^{6.5}$ and $10^{1.5} \lesssim \rho_{\rm g} /\langle \rho_{\rm g} \rangle \lesssim 10^{2.5}$  gas particles in the outskirts of virialized objects currently not observed.
Black dots: currently unobserved intergalactic medium including the warm-hot phase.
Filled white squares gas detectable through OVI absorption lines.
Filled magenta dots gas detectable through OVII absorption lines.
Filled cyan triangles  gas detectable through OVIII absorption lines.
Only absorption lines with $EW$ $\ge \ 10$  $\kms$ have been considered.}
\label{fig:phase}
\end{figure}

Thanks to our WHIM models we can substantiate our previous statement
that observed OVI lines have probed a few per cent
of the missing baryons that reside in the relatively cool regions.
In Fig.~\ref{fig:phase} we show the phase-space diagram of the cosmic baryons at $z \sim 0$
in the hydrodynamical simulation of Borgani \etal (2004).
Small-size points represent gas particles in the simulation.
Some regions of this phase-space diagram have been already probed by observations, like the  X-ray emitting gas in the central regions of virialized objects (the green points with
$T_{\rm g} \gtrsim 
10^{6.5}$ K and $\rho_{\rm g}/\langle \rho_{\rm g} \rangle
\gtrsim 10^{2.5}$),
the diffuse, cold  gas  responsible for the local  Ly $\alpha$ forest (blue points with $T_{\rm g} \le 10^5$ K and
$\rho_{\rm g}/\langle \rho_{\rm g} \rangle \le 10^3$) and the cold, dense gas observed in association to star forming regions within galaxies (yellow points with $T_{\rm g}\le 10^5$ K
and $\rho_{\rm g}/\langle \rho_{\rm g} \rangle \ge 10^3$).
The majority of the remaining baryons are located in regions of the phase space that are currently not probed by observations. Some of them
are expected to be found in the outskirts of the galaxy clusters all the way out to the virial radius (red points with $T_{\rm g}\lesssim 10^{6.5}$ K and $10^{1.5} \lesssim \rho_{\rm g}/\langle \rho_{\rm g} \rangle \lesssim 10^{2.5}$)
but the large majority, which   constitute the bulk of the WHIM, is
to be found in lower density environments

The larger symbols characterize the gas that can be detected through OVI (filled white squares),
OVII (filled magenta dots) or OVIII (filled cyan triangles) assuming model $B2$. Only lines
with  $EW$ $\ge  10 \ \kms$ are considered in the plot.
Different ions preferentially probe different regions of the phase space with a non negligible overlap,
spanning a large range
of WHIM physical conditions. As anticipated, OVI is a signpost for the cooler regions
of the WHIM while, on the opposite, the OVIII lines typically flag hotter environments.
The OVII absorbers span a very large range in temperature, although those with $T_{\rm g} < 10^5$
K are typically associated with weak lines.

The position of the OVI, OVII and OVIII symbols in the phase space is not determined precisely
since the density and the temperature of the  gas associated to the absorption line
are not uniquely defined.
In this work we estimate temperature and density of the absorbing material as in Chen \etal (2003).
First we define  the temperature and density in each spectral bin as
\begin{equation}
\rho_{\rm g}(u)=\frac{1}{\tau(u)} \int \frac{d\tau}{dy} \rho_{\rm g}(y) dy
\  ;  \ \ \ \ \ \
T_{\rm g}(u)=\frac{1}{\tau(u)} \int
 \frac{d\tau}{dy} T_{\rm g}(y) dy
 \label{eq:t}
 \end{equation}
and then we obtain the corresponding quantities for the absorbers by
averaging over the  bins in the line redshift interval $[\underline{z},\overline{z}]$.
This is the definition that we have used to plot the symbols in Fig.~\ref{fig:phase}.

We have also considered an alternative estimator
in which temperature and density are weighted by the optical depth of the line:
\begin{equation}
\hat{\rho}_{\rm g}=\frac{\int^{\overline{z}}_{\underline{z}} \tau(y) \rho_{\rm g}(y) dy}{\int_{\underline{z}}^{\overline{z}} \tau(y)  dy}
\  ;  \ \ \ \ \ \
\hat{{\rm T}}_{\rm g}=\frac{\int^{\overline{z}}_{\underline{z}} \tau(y) {\rm T_{\rm g}}(y) dy}{\int_{\underline{z}}^{\overline{z}} \tau(y)  dy}.
\label{eq:2}
\end{equation}
Using Eq.~\ref{eq:2} rather
Eq.~\ref{eq:t} does not introduce 
systematic differences, just a random scatter of
$\sim \ 15 \%$ in both temperature and density.

Another possible source of ambiguity in estimating the temperature ad the density of 
the gas  is represented by those absorption systems  characterized by two
or more ions, each one with its own opacity. 
In this case, the  redshift interval $[\underline{z},\overline{z}]$ depends 
on the ion considered and so will be the estimated temperature and density of the 
absorbing gas. To assess the importance of this effect we have considered 
all mock absorption systems featuring  any two lines
among OVI, OVII or OVIII and found  $\sim 20$ \%  scatter
in temperature and density  computed in the redshift ranges of the 
different lines.

\subsection{Correlation between absorbers}

\label{sec:linecorrelation}

\begin{figure}
\centering
\epsfig{file=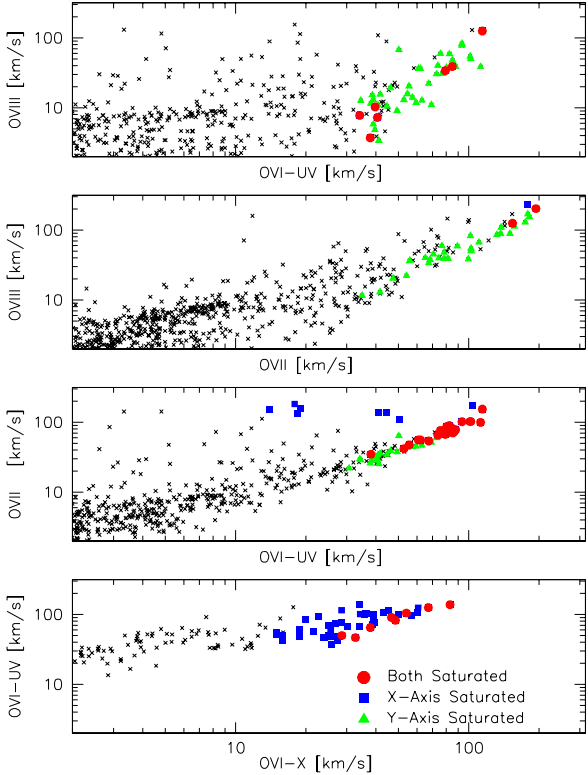,width=0.62 \textwidth}
\caption{Equivalent width correlation for oxygen ions. From top to bottom: OVIII vs. OVI-UV,
OVIII vs. OVII, OVII vs. OVI-UV, OVI-UV vs. OVI-X. Black crosses indicate unsaturated lines.
Filled green triangles: saturated line for the absorber on the X-axis.
Filled blue squares: saturated line for the absorber on the Y-axis.
Filled red circles:  both ion lines saturated. }
\label{fig:correlation}
\end{figure}

Figure~\ref{fig:correlation} shows the predicted correlation between the equivalent widths of the oxygen lines associated to the same absorber 
in the mock spectra of model $B2$. We only consider absorbers in which all  lines
are detected above threshold. To avoid ambiguity we measure the line $EW $ in the velocity range used to identify the OVII line.
In the plots crosses  indicate lines that are not saturated (i.e. for which the transmitted
flux is above zero).
Green triangles (blue squares) indicate absorbers in which the ion line on the X (Y) axis is saturated.
Red dots indicate that both ion lines are saturated.
The bottom panel shows the correlation between the equivalent widths of the OVI UV-doublet
and  the OVI X-ray line. For optically thin lines there is a simple, deterministic
relation between the $EW_{\rm s}$ of the two lines. The scatter in the plot arises from
having measured the $EW$ of the lines in the velocity range appropriate for the OVII line.
Equivalent widths of saturated UV-absorbers deviate from the expected correlation,
causing the flattening at large $EW_{\rm s}$.
There is a small range of temperatures in which both OVII and OVI have high fractional abundance.
In this interval both species trace the underlying baryon density, causing the
tight correlation seen in the third panel of Fig~\ref{fig:correlation}. Strong OVII lines, however, have
a large range of OVI $EW_{\rm s}$. Part of this spread is due to line saturation and
part of it is due to the fact that in warm absorbers the OVI fraction is very low. As pointed out by
Chen \etal (2003) who performed the same correlation analysis, this fact suggests that, at least for
the  stronger lines, follow-up X-ray observations of known OVI absorbers are a promising way to find 
OVII absorption lines  while the opposite strategy is less efficient. Similar considerations apply to OVII and OVIII lines. The scatter in the relation, however, is smaller
than in the previous case since these ions have large fractional abundance over a large range
of temperatures, a fact that is  clearly illustrated in the phase-space diagram of Fig.\ref{fig:phase}.
Finally, there is little or no correlation between  OVI and OVIII except for strong lines.
In this case the large scatter reflects the fact that,  for a fixed gas density, the temperature of
these two species can span a three order of magnitudes range. As a result, OVI lines
are not able to   efficiently flag the presence of   OVIII absorbers and vice versa.

\section{Detecting the WHIM through absorption}
\label{sec:whimetections}

In this Section we explore the possibility of detecting WHIM in the absorption spectra
of X-ray bright objects. We will focus on GRBs, showing that these transient
sources offer several advantages over  bright AGN like quasars and
Blazars. Then we  use our WHIM models to estimate the number of
detections that one may expect from  a {\small  XENIA}/{\small EDGE}-like satellite mission.

\subsection{GRBs as background sources}
\label{sec:grb}

Unambiguous WHIM detections require X-ray background sources that are bright, distant 
and common enough to provide a good statistical sample of WHIM lines.
The latter 
requirement excludes rare objects, like the Blazars during their active phase. Bright 
AGN are quite common but nearby, hence probing relatively short lines of sight. 
Cross-correlating the VERONCAT catalog of Quasars and AGN (V\'{e}ron-Cetty and 
V\'{e}ron 2006) with the ROSAT All-sky Survey Bright Catalogs (1RXS, Voges \etal 
1999), Conciatore \etal (2008) found only 16 objects (mostly BLLac) at $z > 0.3$ with 
a 200 ks integrated flux (fluence) in the [0.3-2] keV range above $10^{-6} \ {\rm 
erg} \ {\rm cm}^{-2}$, bright enough to detect WHIM lines
The mean redshift of these objects is $\sim 0.5$, to be compared with the 
redshift distribution of the Warm-Hot gas (see e.g. Fig. 1 of Cen \& Ostriker 2006) 
that, at  $z\simeq 2$ still accounts for $\sim 20 \%$ of the barionic mass in the universe.
GRBs look more promising in all respects. First of all their X-ray 
afterglows provide a fluence in the [0.3-2] keV similar to that of
bright AGN, collected however in only 50 
ks rather than 200 ks. The only disadvantage is represented by their transient nature 
and by the fact that, since they emit a considerable fraction of the soft X-ray photons 
soon after the prompt, they require an instrument capable of fast re-pointing.

In this work we will focus on GRBs as background sources since the Wide Field 
Monitor proposed for the {\small EDGE/XENIA} missions is especially designed to 
trigger on and localize GRBs and X-ray Flashes and the satellite can re-point the 
source in 60 s. Spectra would be taken by the Wide Field Spectrometer in the 0.2 - 
2.2 keV band by integrating the X-ray flux between 60 s and 50 ks. 
To convert counts to fluxes, or fluxes from one band to another, and to correct for
absorption in our Galaxy and in the host galaxy
we assume a typical power-law GRB spectrum with 
photon index $\Gamma=2$, absorbed by ${\rm N}_{\rm H}{\rm (Galaxy)}=2 \times 10^{20} \ 
{\rm cm}^{-2}$ and by ${\rm N}_{\rm H}{\rm (Source)}=3.2 \times 10^{21} \ {\rm 
cm}^{-2}$.

Given that {\small EDGE/XENIA} is expected to have orbit and slewing capabilities 
similar to those of {\small Swift}, a reliable way to estimate the expected afterglow 
fluence distribution in a given energy band is to use the total number of photons 
actually detected for each GRB by {\small Swift/XRT} in that energy band and 
converting counts to physical units by adopting the best fit spectral model of each 
GRB. In this way, all the effects that may affect the fluence measured by the X-ray 
telescope (e.g.: different times to go on target, Earth occultation statistics and 
South Atlantic Anomaly passage for a satellite in Low Earth Orbit, combined with the 
different shape and duration of both X--ray prompt and afterglow emissions, X--ray 
flares, long duration prompt emission, possibility of triggering on a precursor and 
thus being on target during the main phase of prompt emission, etc.) are 
automatically taken into account.

In the light of these considerations, we have considered, for a sample of 187 GRBs, the 
total number of counts measured directly by {\small Swift/XRT} in the [0.3-1] keV 
band during follow-up observations.
We stress that the fluence 
in the narrow [0.3-1] keV energy band allows to trace  well
the monochromatic 0.5 keV 
fluence which is more relevant to assess the possibility of detecting WHIM lines.
In fact,  the errors caused by the possible scatter of the GRB population around the 
typical spectrum, which is used to convert the observed counts in a 0.5 keV fluence 
(see column 1 in Table 3), are minimized by using the narrower [0.3-1] keV energy 
band.

Based on this approach, we estimated that the afterglows with [0.3-1 keV] total counts 
corresponding to a monochromatic  fluence 0.5 keV
higher than 8.8, 22 and 43$\times$10$^{-7}$ erg cm$^{-2}$ keV$^{-1}$ 
are, respectively, 9\%, 3\% and 1\% of the total.
These fluence thresholds are listed for reference in the first 
column of Table~3. In the second column we also list the corresponding value of the 
[0.3-10] keV unabsorbed (i.e., corrected for the absorption of the Galactic and 
the intrinsic hydrogen-equivalent column density) fluence, derived from the [0.3-1] keV counts by assuming 
the typical GRB afterglow spectrum specified above.
 Taking into account the 
{\small Swift} detection rate of $\sim$100 GRBs per year and the larger solid angle 
of the WFM  on board of {\small EDGE/XENIA}
with respect to the {\small SWIFT/BAT} monitor (a factor $\sim$2 for 
bursts with prompt  fluence 
of about 1$\times$10$^{-6}$ erg cm$^{-2}$, and an additional factor $\sim$3 
for larger values of the fluence, see Piro \etal (2008) for details), one expects 
to observe about 18, 8 and  3 such events per year, respectively.

\begin{figure}
\centering
\epsfig{file=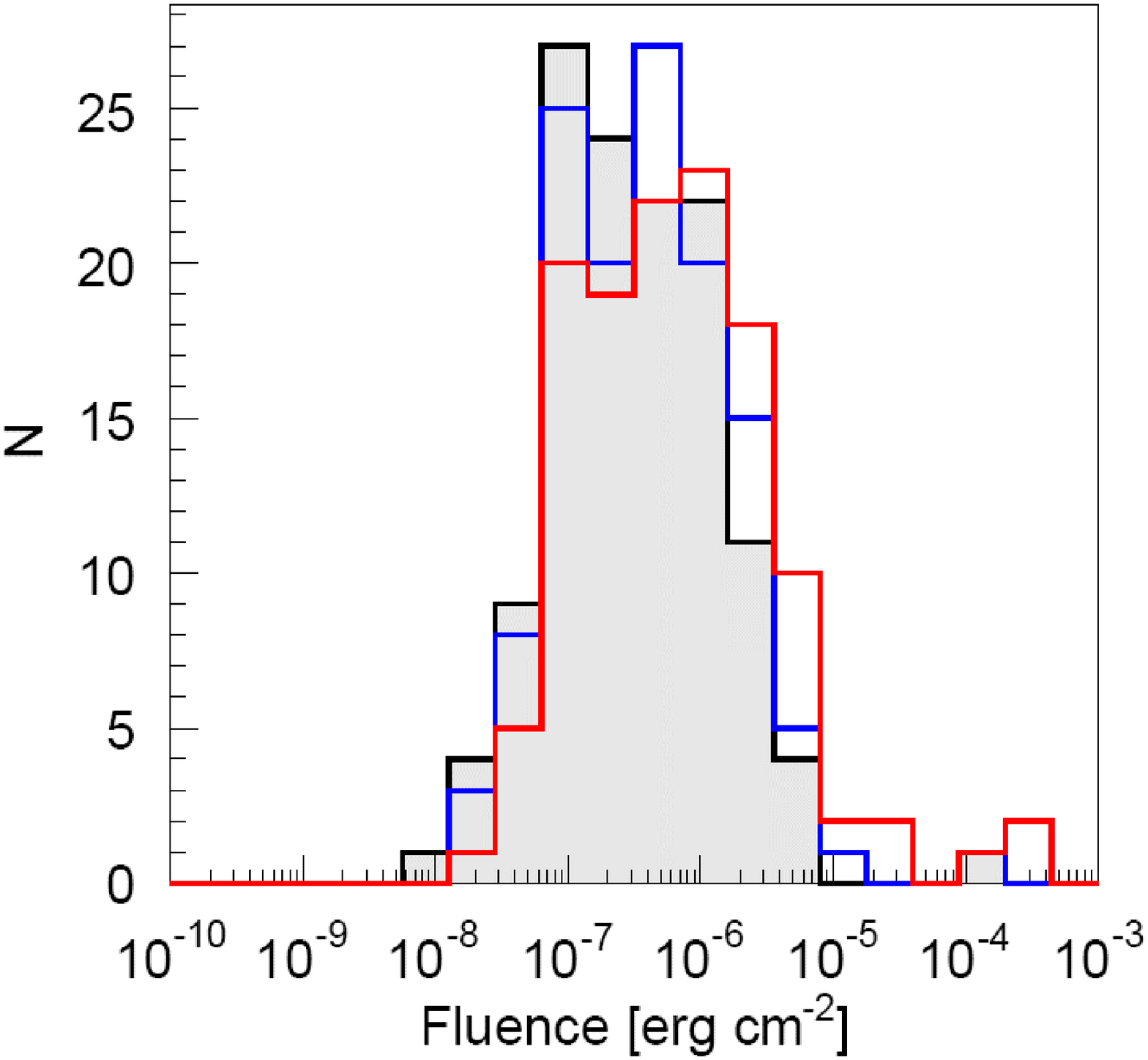, angle=0, width=0.5 \textwidth}
\caption{X-ray [0.3-10] keV fluence distribution for 125 GRBs observed by {\small Swift}
with observation
start time $<$ 1000 s. The three histograms assume different start times and uninterrupted integration intervals. Red curve: 60 s - 60 ks. Blue curve: 120 s - 60 ks. Black curve limiting the shaded area: 180 s - 60 ks.}
\label{fig:swiftaft}
\end{figure}

These numbers represent a conservative estimate based on current {\small 
Swift} capabilities. They can be increased in two ways. First of all, reducing the 
slewing time from the average 120 s required by {\small Swift} to 60 s would 
increase the count fluence by a factor $\sim$2. The amplitude of this effect is 
quantified by the change in the distribution of the unabsorbed fluence in the 
[0.3-10] keV shown in Fig.~\ref{fig:swiftaft}, where different observation start times 
from the burst onset are assumed. These fluences have been derived 
from a sample of 125  {\small Swift/XRT} light curves, assuming uninterrupted 
observations (the observed flux in {\small Swift}, due to the low earth orbit, is 
basically half of that of an uninterrupted observation), and converting counts to 
physical units by adopting the best fit spectral model of each GRB. The blue 
histogram shows the case of a start time from the burst onset of 120 s, equal to 
the average {\small Swift} case. Using a satellite like {\small EDGE/XENIA}, for 
which the slewing time could be reduced to 60 s, would shift the distribution 
toward significantly larger values of the fluence, as shown by the red histogram.
In that histogram there  are 3 events exceeding 10$^{-4}$ erg cm$^{-2}$.
Two of them are due to extrapolation of the initial very steep decay.  The third one is
a really bright but  very peculiar event (GRB 060218) with a very long ($>$ 35 min) prompt
emission which was seen by {\small Swift/XRT} too.

\begin{figure}
\centering
\epsfig{file=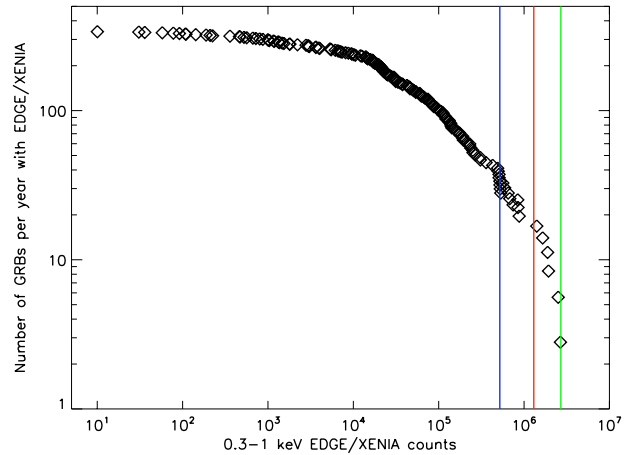, angle=90, width=0.45\textwidth}
\caption{Cumulative distribution of the GRB X--ray afterglows
detected per year by a satellite like {\small EDGE/XENIA}
as a function the unabsorbed fluence in the 
[0.3-1] keV band. 
Vertical lines are drawn at the three different 
fluence thresholds of 2, 5 and 10 $\times$10$^{-6}$ erg cm$^{-2}$
in the [0.3-10] keV band. These have been obtained assuming 
the typical GRB afterglow  spectrum and column density described in the text.}
\label{fig:edgecounts}
\end{figure}

The second improvement is represented by the possibility of detecting and following up 
of the softest GRB, i.e. the X--Ray Rich (XRR) GRBs and the X-Ray Flashes (XRF) 
whose existence has been revealed by {\small BeppoSAX} (Heise \etal 2001). Their 
abundance turned out to be comparable to that of classical GRBs. In addition, the 
observations by {\small BeppoSAX} (D'alessio \etal 2006) recently complemented 
by {\small Swift XRT} (Gendre \etal 2007) showed that, despite of a less 
energetic prompt gamma-ray emission, the X-ray afterglow flux of XRFs and XRRs is 
similar to that of GRBs. As a consequence, detecting these populations of objects 
would significantly increase (almost by a factor of two)
 the number of bright afterglows that can be used to study the WHIM.
 Indeed XRFs constitute  about 30\% of the  {\small HETEII} sample, while the combined
XRF+XRR sample amounts to 70\% of the total (Sakamoto \etal 2005). For reference, 
XRFs constitute only about 10\% of the {\small Swift} sample (Gendre 2007). This 
difference is due to the higher lower energy threshold (15 keV) of the BAT trigger 
instrument, compared to {\small HETEII WXM} (2 keV) or to {\small BeppoSAX WFC}. 
Adopting a 
low energy threshold of 8-10 keV, which represents the baseline for the monitor 
planned for {\small EDGE/XENIA}, is expected to increase by $\sim$30 \% the number 
of detections. With a low energy threshold of 5 keV the increase would be of $\sim 
70$\%.

The cumulative distribution of GRBs that {\small EDGE/XENIA} is expected to observe in 
one year as a function of the counts measured in the [0.3-1] keV band is shown in 
Fig.~\ref{fig:edgecounts}  (see the EDGE science goal document available at 
http://projects.iasf-roma.inaf.it/ for further details). 
Assuming a typical GRB spectrum, we draw three vertical 
lines at the same reference 0.5 keV monochromatic fluences of
8.8, 22 and 43$\times$10$^{-7}$ erg cm$^{-2}$ keV$^{-1}$
 considered above, that correspond to the [0.3-10] keV unabsorbed
fluence values listed in column 2 of Table 3. The number of expected detections 
above these reference fluence values is 48, 18 and 8 respectively.

In column 3 of Table~3, we report the number of GRBs 
that {\small  EDGE/XENIA } is expected
to detect above a given fluence threshold, computed by averaging
between the pessimistic case, obtained by simply scaling the {\small Swift}
results, and the optimistic case, which accounts for the optimization 
in the slewing capabilities and the possibility of detecting XRFs and XRRs in addition 
to classical GRBs.  

As it can be seen, {\small EDGE/XENIA }  is expected to detect in 3 years  $\sim 100$ GRBs 
above the fluence threshold indicated in Table~3 compared to the 16
distant AGN potentially useful for WHIM detection, if observed for 200-ks each.
In fact, as we have stressed already,
the situation is even more favorable for the WHIM detection since
the average redshift of GRBs is significantly larger that
AGN and the probability of line detections is expected to increase accordingly.

\begin{deluxetable}{cccccccc}
\tablecaption{GRBs and WHIM line detection}
\label{tab:tab3}
\tablehead{
\colhead{F\tiny{$_{\mbox{0.5}}$}}  &
\colhead{F\tiny{$_{\mbox{0.3-10}} $}}  &
\colhead{ \#\tiny{$_{\mbox{\rm GRBs}}$}}  &
\colhead{ EW\tiny{$_{\mbox{\rm OVII}}$}} &
\colhead{ \#\tiny{$_{\mbox{\rm OVII}}$ }} &
\colhead{EW\tiny{$_{\mbox{\rm OVIII}}$}} &
\colhead{\#\tiny{$_{\mbox{\rm OVII+OVIII}}$ }}  
}
\startdata
$8.8 \ 10^{-7}$ & $ 2 \ 10^{-6} $ & 33 & 0.28 &  58 & 0.19 & 37  \\  
$2.2 \ 10^{-6}$ & $ 5 \ 10^{-6} $ & 13 & 0.18 &  47 & 0.12 & 29 \\  
$4.3 \ 10^{-6}$  & $ 1 \ 10^{-5} $ &   6 & 0.12 &  32 & 0.08 & 19 \\  
\enddata
\tablenotetext{}{\tablenotetext{}{Column 1: Ion. Column 2: Wavelength [eV]. Column 3: Oscillator Strength.}Column 1: Fluence at 0.5 keV [erg cm$^{-2}$ keV$^{-1}$].
Column 2: [0.3-10] keV Unabsorbed fluence [erg cm$^{-2}$] collected 
during the GRB follow-up observations.
Column 3: Number of GRBs per year above fluence in a field of view of 2.5 sr.
Column 4: Minimum $EW$ (in eV) for a 4.5$\sigma$ OVII line detection in the $B2$ model.
Column 5: Number of  4.5$\sigma$ OVII line detections per year in the $B2$ model.
Column 6: Minimum $EW$ (in eV) for a 3$\sigma$ OVIII line detection in the $B2$ model.
Column 7: Number of  simultaneous OVII and OVIII line detections per year
with significance $\ge5 \  \sigma$ in the $B2$ model.
}
\end{deluxetable}

\subsection{Probability of OVII detection}
\label{sec:onedetection}

In this Section we quantify the detectability of WHIM absorbers by applying the treatment and formalism
of Sarazin (1989) to transient sources. We first focus on the problem of detecting the OVII
line, which, as stressed before, is usually  the strongest absorption feature for the warm-hot
intergalactic  gas.

\begin{figure}
\centering
\epsfig{file=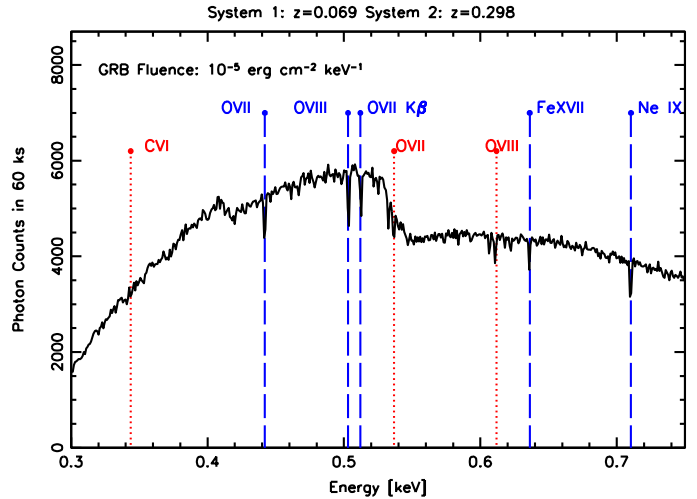,width=0.53\textwidth}
\caption{Mock X-ray absorption spectrum of a typical GRB  afterglow at $z=2.6$ convolved with the
{\small EDGE/XENIA} response matrix. Dotted vertical lines mark the most prominent absorption lines produced by a WHIM filament at $z=0.069$. A second absorption system
is also present along the LOS at $z=0.298$. Its absorption lines are flagged with the vertical dashed lines.
The ions responsible for the absorption features are indicated in correspondence of each line.
The fluence at 0.5 keV of the afterglow is also shown.}
\label{fig:convolved}
\end{figure}

In Fig.~\ref{fig:convolved} we show one realistic mock absorption spectrum of a GRB  X-ray
afterglow  convolved with the \small{EDGE/XENIA WFS } response matrix. Two WHIM absorbers
at two different redshifts (indicated in the plot) are present. In both cases
the OVII absorption
line stands out prominently along  with the OVIII line. 
The nearest absorber at $z=0.069$ also shows
a weak CVI line. 
The absorber at $z=0.298$ instead, is characterized by
considerably stronger  absorption features including the FeXVII, NeIX and  the OVII 1s-3p lines.

In order to detect an absorption line with statistical significance  $\sigma$, the detector must collect,
within the instrument energy resolution, $\Delta E$,  a number of continuum photons
$N_{photons}\magcir \left(\frac{S}{N}\right)^2\left(\frac{EW}
{\Delta E}\right)^{-2}$.
The corresponding minimum equivalent width $EW_{\rm min}$
for a given line fluence $F$ is:
\bea
\nonumber
\left(\frac{EW_{\rm min}}{100\, \rm{km/s}}\right) \magcir \left(\frac{F}{1.3 \times 10^{-7} \rm{erg\, cm}^{-2}\,
\rm{keV}^{-1}}\right)^{-0.5}  \times \\
\left(\frac{E}{1\,\rm{keV}}\right)^{-0.5}\left(\frac{\sigma}{3}\right)
\left(\frac{\Delta E}{10\, \rm{eV}}\right)^{0.5}
\left(\frac{A}{10000\,\rm{cm}^2}\right)^{-0.5} \, ,
\label{fluence}
\eea
where $A$ is the effective collecting area.

In column 4 of Table~3
we list the minimum equivalent width required to detect
the  OVII  line with a significance of 4.5 $\sigma$ assuming an energy resolution
$\Delta E=$  3 eV and a collecting area  $A=$ 1000 cm$^2$ (i.e. matching the baseline configuration
of the Wide Field spectrometer of {\small EDGE/XENIA}) in the X-ray spectra of a GRB
with a 0.5 keV fluence 
indicated  in column 2 of the same Table.
These estimates have been obtained by setting $E=0.574/(1+0.32)=0.435$ keV in eq.~\ref{fluence}
i.e. by assuming that all absorbers are at $z=0.32$ which is the 
average redshift of mock absorbers with $EW>0.08$ eV  in the $B2$ model
The reason for considering 4.5 $\sigma$  significance is that it guarantees 
that when  two lines 
 (typically OVII and OVIII) can be detected, the underlying absorber is detected  
with   5 $\sigma$ significance, as will become clear  in the next Section. 
The corresponding  number of  OVII line detections with significance $\ge \ 4.5 \ \sigma$
is listed in column 5. It was  obtained by counting the number of OVII absorbers 
with $EW > EW_{\rm min}$ (in column 4) 
in the mock spectra of model $B2$ along a number of lines-of-sight equal to that 
of the expected GRBs listed in column 3.
We recall that the average number of OVII lines above $EW$ is shown in  the left panel of Figure~\ref{fig:ion1}, in which we can appreciate the differences among the various WHIM models' predictions.
Using models $M$ or  $B1$ instead of model $B2$ which we regard
as better physically motivated, significantly reduces the number of expected detections. 
The exact magnitude of the effect depends on the $EW$  of the line and on  the GRB fluence.
 We estimate that, even with  the most conservative model, a satellite like
  {\small EDGE/XENIA} should be able to detect about 25 OVII lines per year.

We point out that, in all model explored,
the number of expected OVII detections listed in Table 3 is biased low
since  in our mock spectra we have only considered OVII lines at $z<1$ ($z<0.5$ in model
$M$) while, as we have already pointed out, significant amount of WHIM gas 
are expected out to significantly higher redshifts.
In addition, all these estimates should be boosted up by $\sim 40$ \% if the energy resolution 
of the spectrograph could improve to 1 eV, which constitutes the goal for the {\small WFS}.

These estimates for a single OVII  line detection, however, must be regarded as  hypothetical
since  with no \emph{a-priori} knowledge of  the absorber's redshift there is no efficient way of preventing
contamination by spurious absorption features or confusion with absorption lines form other ions.
Of the two effects,  confusion should be regarded as the minor one
given the paucity of expected non-Galactic absorption lines in the energy range relevant for the WHIM.
An effective procedure to  minimize confusion is suggested by the
analysis of our mock spectra. The idea is to identify the strongest  absorption feature with the  OVII  line (which is often the case), look for all other possible ion lines at the same redshift as the OVII line then move to the next strongest  line and so on. 
The second effect, contamination by Poisson noise,  should be dominant. However, its 
importance decreases with  the $EW$ of the OVII line. To quantify the amount of contamination we have produced a number of mock GRB spectra  consisting in  Monte Carlo realizations of  power-law spectra with a specified ${\rm N}_{\rm H}$ absorption, convolved with the  {\small EDGE/XENIA}  response matrix, all of them  with a line fluence  F$_{0.5}$ in the range $ [0.9-4.3] \times 10^{-6} $ erg cm$^{-2}$ keV$^{-1}$.
These mock spectra have been "observed" to compute the probability of fake detections resulting from Poisson fluctuations as a function of the $EW$ of the spurious line.
By searching for OVII lines in the energy range $[0.287-0.574]$ keV, corresponding to a redshift range
$z= \ [0-1]$, we computed the minimum $EW$ required to guarantee a contamination level below 1\%.
The resulting $EW_{\rm s}$ are the same as those listed in column 4 of Table~3, i.e. they  correspond to the  minimum $EW_{\rm s}$ required to guarantee a 4.5 $\sigma$ detection of the OVII line. In other words, a  4.5 $\sigma$ detection of a putative OVII line guarantees that contamination by Poisson noise is below  1 \%.

This exercise illustrates that the simple search for a single OVII line in the GRB spectra observed by {\small  EDGE/XENIA}  would  provide 25-80 line detections per year, where the lower bound is obtained by considering the most pessimistic among  our WHIM models and the upper bound assumes a goal energy resolution of 1eV.
We stress again that out of this sample, we expect at most 1 \% to be spurious.

Finally, we note that the smallest minimum $EW$  in Table~3
for a joint  OVII and OVIII lines detection, which we investigate in the next Section,
 is 0.08 eV. These values require a control of the 
systematic effects, in particular of the energy width of instrumental bin 
and the level of the continuum, at  $\sim$ 2 \% or better, corresponding to a  systematics of 
0.06 eV for 3 eV resolution ($\sim$ 8 \% for a goal energy resolution of 1eV).

\subsection{Probability of multiple line detections}
\label{sec:moredetections}

X-ray spectroscopy allows in principle to characterize the physical state of the WHIM and measure its metal content. Unfortunately, the current energy resolution of the microcalorimeters in the soft X-ray band does not allow to resolve the WHIM lines and estimate the gas temperature directly from the line width. Indeed, typical line widths of detectable WHIM absorbers in our mock spectra are in the
range 0.1-0.2 eV, well below the goal energy resolution of 1 eV of TES microcalorimeters.
Nevertheless, as we shall discuss in the next Section, we can still use unresolved lines to
constrain the physical
state of the absorbing gas, provided that we can detect the absorption line of some other ion  in addition to OVII.

To asses the possibility of multiple line detection we first consider  the OVIII line which often constitutes  the second strongest absorption feature in the mock spectra. Having already identified the redshift of the absorber from the OVII line makes the detection of the OVIII line comparatively easier since one can now be happy with a  3$\sigma$ rather than a 4.5$\sigma$ detection. The minimum $EW$ for such detection can be obtained from  Eq.~\ref{fluence} and is listed in column 6 of Table~3 
as a function of the GRB fluence. 
Like in the previous case of single OVII line detection, we have only considered OVIII absorption 
systems with  $z\le 1$ and evaluated the minimum equivalent width of the line at $z=0.32$, i.e. using $E=0.495$ keV in Eq.~\ref{fluence}.

We find that  $\sim 50$ \% of all $3\sigma$ OVIII lines  are associated to a $4.5\sigma$ 
OVII line, i.e. the cumulative distribution function of OVIII and OVII line pairs lies a factor of two  below 
that of the OVIII line alone, plotted in the right panel of  Figure~\ref{fig:ion1}. 
The average ratio between $EW_{\rm min}^{\rm OVIII}$ and $EW_{\rm min}^{\rm OVII}$ is $\sim 0.7$, that is to 
detect the OVIII line  associated to a 4.5 $\sigma$-detected OVII line the following condition must be satisfied
$EW_{\rm OVIII} \  > \ 0.7 \ EW_{\rm OVII}$. Moreover, using Monte Carlo simulations we have verified 
that a joint OVII    4.5 $\sigma$  and OVIII  3 $\sigma$ detection 
corresponds to an overall $\sim 5$ $\sigma$ detection
of the underlying absorption system, further reducing the probability 
of spurious lines contamination.

Since the bulk of the WHIM is not in collisional ionization equilibrium,
the $EW$ ratio of two lines depends on both the temperature and the density of the absorbing
gas. Therefore, the simultaneous observation of the  OVII and OVIII lines alone
cannot constrain the physical state of the WHIM. Instead, as we will show in the next Section,
a third line, preferably of the same element, must be detected as well. 
To estimate the number of  absorbers featuring at least three lines we looked for all mock absorption  systems in which an additional   3 $\sigma$  line is detected in association with  a 
4.5 $\sigma$ OVII line and 3 $\sigma$ OVIII line.
Table~4  shows the expected number of such simultaneous detections per year as a function of the GRB fluence.
We did not include OVII 1s-3p in the Table since its detection would only increase the statistical significance of the absorption system but give no insight on its physical state  (the 
$EW$ ratio of the  OVII 1s-3p  and  OVII 1s-2p lines only depends on their relative oscillator strength).

Detecting 3 lines of the same elements, like OVI, OVII and OVIII, would allow to probe the physics of the WHIM directly. Unfortunately, as shown in Table~4, the X-ray  OVI line is very weak and the number of expected 
multiple oxygen  lines detection per year is very small. For this purpose one should use the OVI UV line whose detection in the GRB spectra would require  simultaneous UV and X-ray spectroscopy.

Table~4  shows that the best chances to observe multiple line absorption systems would be through the CVI or NeIX lines. In these cases, however, to assess the physical state of the WHIM,
 the relative  abundance of the elements needs to be specified {\it a priori}
Sticking to  these ions
we have found very few cases of multiple detections which does not include both the OVII and the OVIII line. Also, a significant fraction of multiple line detections consists of more than three lines, i.e. the total number of multiple detections expected per year is less than the sum of all entries in the columns of Table~4.

\begin{deluxetable}{ccccccc}
\tablecaption{Multiple WHIM lines detections involving OVII  1s-2p, OVIII
and a third ion}
\label{tab:tabmulti}
\tablehead{
\colhead{\#\tiny{$_{\mbox{0.5}}$}}  &
\colhead{\#\tiny{$_{\mbox{\rm OVI-X }}$}} &
\colhead{\#\tiny{$_{\mbox{\rm CV}}$ }} &
\colhead{\#\tiny{$_{\mbox{\rm CVI}}$ }} &
\colhead{\#\tiny{$_{\mbox{\rm NeIX}}$ }} &
\colhead{\#\tiny{$_{\mbox{\rm FeXVII}}$ }} &
\colhead{\#\tiny{$_{\mbox{\rm MgXI}}$ }}
}  
\startdata
$ 8.8 \ 10^{-7} $ &  4 & 13 &  19 & 23 & 7 & 6 \\ 
$ 2.2 \ 10^{-6} $ &  3 & 12 &  18 & 23 & 6 & 5 \\  
$ 4.3 \ 10^{-6} $ &  3 & 7 &  15 & 15 & 6 & 5 \\  
\enddata
\tablenotetext{}{Column 1: GRB fluence at 0.5 keV [erg cm$^{-2}$ keV$^{-1}$].
Column 2: Detections per year involving OVI 1s-2p.
Column 3: Detections per year involving CV.
Column 4: Detections per year involving CVI.
Column 5: Detections per year involving NeIX.
Column 6: Detections per year involving FeXVII.
Column 7: Detections per year involving MgXI.
}
\end{deluxetable}

\section{Probing the physical state of the WHIM and its cosmological abundance}
\label{sec:whimphysics}

The analysis of X-ray absorption spectra can provide  much more
than a mere WHIM detection. In this Section we address the problem of determining the 
temperature and density of the warm hot gas and  its cosmological density.
To explore the different WHIM models available, all results shown in this Section
were first obtained for model $B1$ and then checked their validity for model $B2$.
Because of the paucity of multiple-line absorption systems, we
did not extend the analysis to the mock spectra of model $M$.

\subsection{Measuring the density and the temperature of the WHIM}
\label{sec:whimrhot}

As already mentioned, the quantitative analysis of the absorption spectra is complicated by the fact that
the X-ray absorption  lines  cannot be resolved by
currently proposed micro-calorimeters.
On the other hand the weak WHIM lines are expected to be optically thin so that a simple
linear relation  holds  between the line
equivalent width and the ion column density:
\beq
EW_{\rm P}(\kms)=\frac{ \pi e^2}{m_e c} \lambda(X^i) f_{X^i} N_{X^i} ,
\label{ewthin}
\eeq
where $X^i$ represents the ion of the element $X$, $N_{X^i} \equiv \int n_{X^i}(y) dy$
its column density, $\lambda(X^i)$  is the wavelength of the transition and
$f_{X^i}$ is the oscillator strength.
From now on we use the subscripts $_{\rm P}$ to
indicate predicted values and  $_{\rm M}$ to indicate the values  measured from the spectra.
In Eq.~\ref{ewthin} $EW_{\rm P}$  is expressed in $\kms$, the same units that we
will use throughout this Section.

Eq.~\ref{ewthin} allows to estimate the  column density of the ion from
the $EW$ of the  absorption line.
However, with no {\it a-priori} information on the size of the absorbing region,
one cannot use the measured $EW$ to estimate the actual density of underlying gas and
assess its cosmological density.
However, some insights on the physical state of the absorber can be obtained from
the $EW$ ratios of at least three different ion lines. In the optically thin limit
the $EW$ ratio between any two lines is:
\beq
\frac{EW_{\rm P}(X^i)}{EW_{\rm P}(Y^j)}=
\frac{ \lambda(X^i)}{\lambda(Y^j)}\frac{f_{X^i}}{f_{Y^j}}\frac{X^i}{Y^j}\frac{A_X}{A_Y} ,
\label{ewratio}
\eeq
where $X_i$ and $Y_j$ represent the two ion fractions
while the ratio  between $A_X$ and $A_Y$ represents
the relative abundance of the element $X$ compared to $Y$. Eq.~\ref{ewratio} has been derived from
Eq.~\ref{ewthin} assuming  that the density of the ions is constant across the absorber
and can be used to infer the density and temperature of the gas upon which
the ionization fractions $X_i$ and $Y_j$ depend. 
Here we apply Eq.~\ref{ewratio} to estimate the density and temperature
of the absorbing systems in our mock spectra, hence using an approach similar to that  of Nicastro \etal (2005) to analyze the physical state of  the  two putative WHIM absorbers along the LOS to Mrk 421. 
For this purpose   we use the same grids  of  hybrid collisional ionization plus photoionization models that we have used to draw the mock spectra from the hydrodynamical simulations. These models are used 
 to predict the ratio ${X^i}/{Y^j}$ which we substitute in Eq.~\ref{ewratio} to obtain $EW_{\rm P}(X^i)/EW_{\rm P}(Y^j)$ as a function of gas density, temperature and ionization background for a given relative metallicity $A_X/A_Y$, assumed {\it a priori}. In a second step, we compare the predicted value 
of    $EW_{\rm P}(X^i)/EW_{\rm P}(Y^j)$ with that 
measured from the mock spectra, $EW_{\rm M}(X^i)/EW_{\rm M}(Y^j)$, to estimate the density and the temperature of the absorbing gas. Unlike Nicastro \etal (2005) we do not attempt to use this approach in combination with the measured $EW$ of the lines to estimate the metallicity of the gas and constrain the hydrogen column density.

\begin {figure}
\centering
\epsfig{file=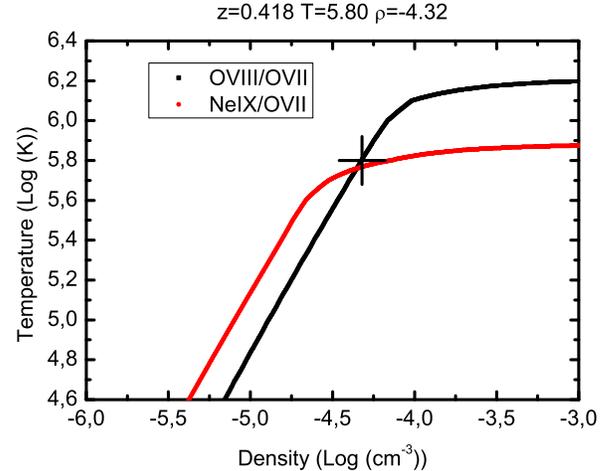,width=0.5 \textwidth}
\caption{
The locii of the minima for the function
$\left[ \frac{EW_{\rm M}(X^i)}{EW_{\rm M}(Y^j)}-\frac{EW_{\rm P}(X^i)}{EW_{\rm P}(Y^j)} \right]^2$
in the $(\rho_{\rm g},T_{\rm g})$ plane.
The two curves refer to
a mock absorber at $z= 0.418$ for which we have detected the OVII, OVIII and NeIX
lines. The cross indicates the expected values.}		
\label{fig:rhotmin}
\end{figure}

In our analysis we only considered absorption systems with unsaturated OVI, OVII, OVIII and NeIX lines with $EW>20 \ \kms$, i.e. strong enough to be detected.
For the OVI  we have also considered weaker lines with $EW > 5 \ \kms $ 
potentially observable in the UV spectra. 
The density and temperature of the gas can be found by minimizing the difference between $EW_{\rm P}(X^i)/EW_{\rm P}(Y^j)$  and $EW_{\rm M}(X^i)/EW_{\rm M}(Y^j)$, and considering at least two independent $EW$ ratios. Figure~\ref{fig:rhotmin} illustrates the outcome of this
minimization procedure. Each curve represents the locus of the minima for the function $\left[ \frac{EW_{\rm M}(X^i)}{EW_{\rm M}(Y^j)}-\frac{EW_{\rm P}(X^i)}{EW_{\rm P}(Y^j)} (\rho_{\rm g},{\rm T}_{\rm g})
 \right]^2$ 
in the phase space, given the  measured  ratio $EW_{\rm M}(X^i)/EW_{\rm M}(Y^j)$.
The two curves refer to the OVIII/OVII and NeIX/OVII ion ratios.
The estimated value for the gas density  and temperature is
found at the  interception between the two curves.
The cross indicated the "true" density and temperature of the absorbing gas.
As already discussed in Section 3.1 these  values 
of $\rho_{\rm g}$ and T$_{\rm g}$
are  somewhat ill-defined since the physical state of the gas  can vary across the absorbing region and
the effective values of its density and temperature depend on ion considered and on the 
estimator adopted to measure its opacity.
This ambiguity can be regarded as an intrinsic uncertainty in the density and temperature of 
the absorbing gas. In Section 3.1 we have used two different estimators (Eq.~\ref{eq:t} and \ref{eq:2})
to quantify this uncertainty and found that the typical difference in both temperature and density 
found when using either estimator is  15-20~\% represented by the size of the cross in 
Fig.~\ref{fig:rhotmin}. To further assess the amplitude of this scatter we have used yet another
estimator in which the temperature and the density of the gas are measured in
correspondence of  the minimum of the absorption line. Using the full battery of the three estimators
confirms that intrinsic uncertainty in the density and the temperature of the absorbing gas 
is 15-20~\%.

The results presented in this Section refer to the 
mock absorbers for which we can measure the
OVII/OVIII and NeIX/OVII $EW$ ratios. Solutions
to the minimization procedure are searched for
in the phase-space region with
$10^{4.5}<T_{\rm g}<10^7 $ K and
$10^{-6} < \rho_{\rm g} < 10^{-3} \ {\rm cm}^{-3}$, corresponding to over-densities
in the range $5<\delta_{\rm g}<5000$, i.e. slightly larger than the canonical WHIM range.
In  $\sim 70$ \% of the cases we have been able to estimate the physical state of the absorbers.
For  the remaining $\sim 30$ \% of the absorbers, we fail to find a common solution, i.e. the 
two curves do not cross in the explored density-temperature range.
We have checked that  if we include the third, non-independent curve relative to
$EW$ ratio of NeIX/OVIII, the resulting additional crossing points  are close to the original one.

To assess the accuracy of our procedures in the $70$ \% successful cases, we have compared
the $\rho_{\rm g}$ and $T_{\rm g}$ estimated through the minimization procedure with
the true values. Random and systematic errors
can be appreciated from scatterplots like those in Figs.~\ref{fig:rr} and  ~\ref{fig:tt} that refer to
the mock absorption systems featuring  detectable OVII, OVIII and NeIX lines.
The estimated gas density turned out to be systematically smaller than the true one by
about 35 \%, similar in size to the random error of $\sim 50$ \%
that we identify with the scatter around the best fitting line in the plots.
The temperature is overestimated by about 35 \%, with a random error of the same amplitude.
Using OVI instead of NeIX leads to similar, although somewhat smaller,
systematic and random errors.

\begin{figure}
\centering
\epsfig{file=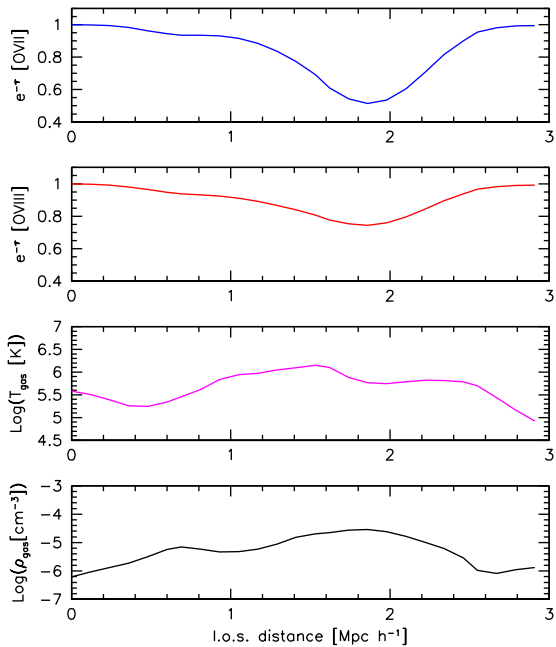,width=0.55\textwidth}
\caption{Example of a typical unsaturated WHIM absorber in our mock spectra
from model $B1$.
{\it From Top to Bottom}: Transmitted flux for OVII, OVIII, temperature and density 
of the gas along the
LOS distance.}
\label{fig:spzoom}
\end{figure}

These errors may originate either from the presence of strong inhomogeneities within the 
absorbing gas or  from the breakdown of the optically thin approximation. 
To check the validity of the first hypothesis we have examined the behavior of the temperature and the density within the redshift range of the lines corresponding to the WHIM absorbers. 
Fig.~\ref{fig:spzoom} is analogous to Fig. \ref{fig:sp1} and shows a typical unsaturated mock 
absorber. The temperature and the density of the gas are remarkably constant across the redshift
range of the OVII and OVIII lines. This result is at variance with that found by  Kawahara \etal (2006). The authors found large deviation from homogeneities in the absorbing gas, as clearly illustrated 
in  Fig. 15 of their paper. This discrepancy can be understood by noticing that 
Kawahara \etal (2006) absorbers are typically
found in regions of higher density and temperature than ours, i.e. within environments 
in which density and temperature
vary change significantly  across a $\sim$ Mpc scale. Such a difference
probably derives from the fact that the OVII ionization fraction and the overall 
gas metallicity in Kawahara \etal (2006) WHIM model are smaller than in our case.

\begin {figure}
\centering
\epsfig{file=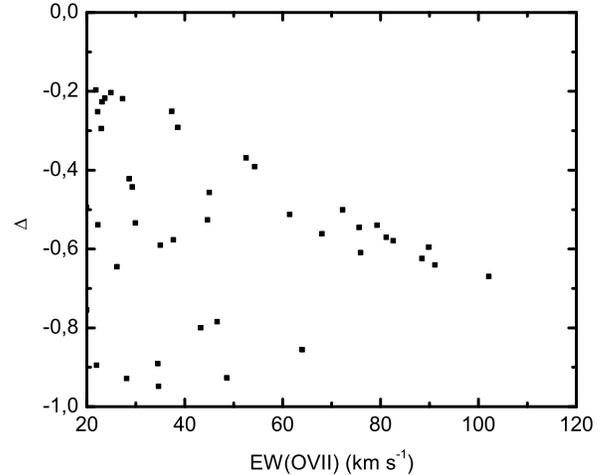,width=0.5\textwidth}
\caption{
Fractional difference $\Delta$, between
$\frac{EW_{\rm P}(NeIX)}{EW_{\rm P}(OVII)}$ and
$\frac{EW_{\rm M}(Ne IX)}{EW_{\rm M}(OVII)} $ as a function of
$EW_{\rm M}(OVII)$.
Absorbers with unsaturated lines with
 $EW>20 \  \kms$  are shown with black, filled dots.
 }
\label{fig:correction}
\end{figure}

\begin {figure}
\centering
\epsfig{file=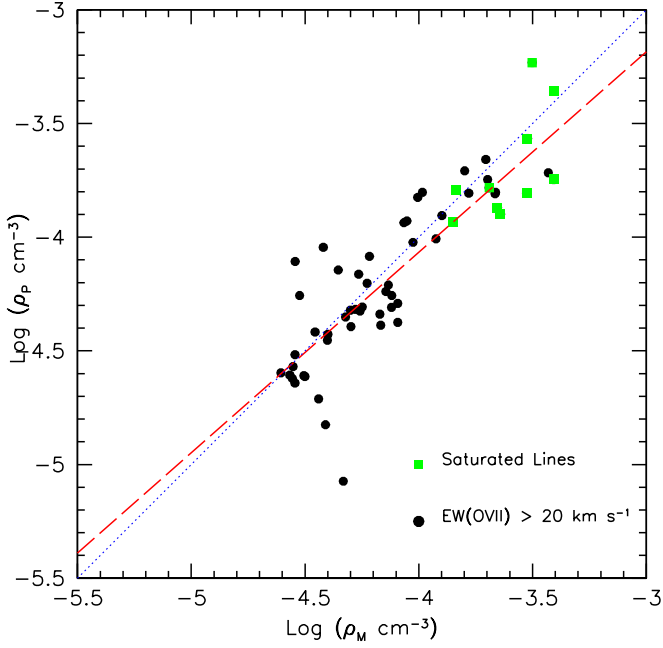,width=0.5\textwidth}
\caption{Scatterplot of the predicted gas density, $\rho_{\rm P}$ vs. measured gas density
$\rho_{\rm M}$ after correction for systematic errors.
Filled dots refer to all absorbers with $EW(OVII) > 20 \ \kms$
and filled, green squares to absorbers which are saturated in any of the lines.
The dashed red line represents the best fit to the unsaturated points.}
\label{fig:rr}
\end{figure}

\begin {figure}
\centering
\epsfig{file=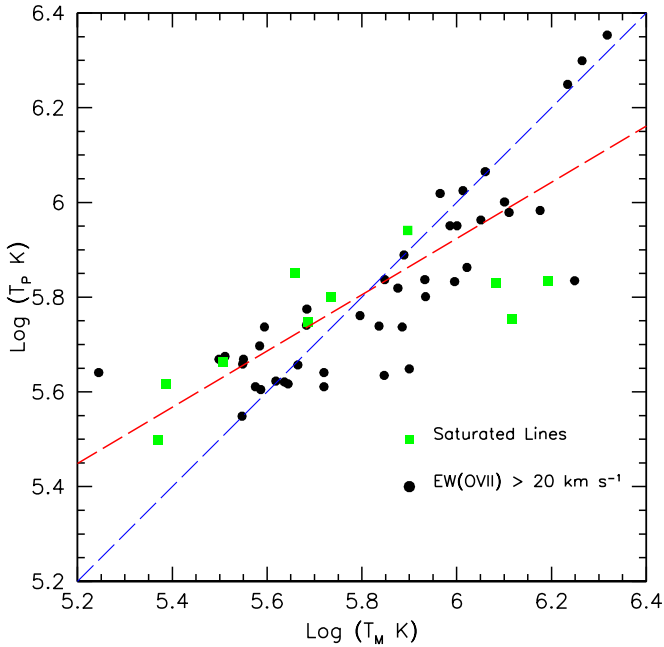,width=0.5\textwidth}
\caption{Scatterplot of the predicted gas temperature, $T_{\rm P}$ vs. measured gas temperature
$T_{\rm M}$ after correction for systematic errors.
Filled dots refer to all absorbers with $EW(OVII) > 20 \  \kms$
and filled, green squares to absorbers which are saturated in any of the lines. 
The dashed red line represents the best fit to the unsaturated points.
 }
\label{fig:tt}
\end{figure}

To test whether the systematic errors in the estimate of  $\rho_{\rm g}$ and $T_{\rm g}$
can be attributed to departures from the optically-thin approximation we have computed 
fractional difference between the measured and the predicted $EW$ ratio
computed using the true values of density and temperature,
$\Delta= \left[
\frac{EW_{\rm P}(NeIX)EW_{\rm M}(OVII)}{EW_{\rm P}(OVII)EW_{\rm M}(Ne IX)} -1\right]$.
 The breakdown of the optically 
thin approximation leads to overestimate the equivalent width of the line,
$EW_{\rm P}>EW_{\rm M}$. This mismatch is more serious for OVII, which is the strongest line, 
and should increase with the strength of the line. As a consequence we expect a negative
$\Delta$ with an amplitude that
 increases with $EW_{\rm M}(OVII)$.
This effect can indeed be appreciated from 
Fig.~\ref{fig:correction} in which, superimposed to 
the scatter,  there is a clear trend for $\Delta$ to decrease with $EW_{\rm M}(OVII)$.
The  scatter is particularly large for weak lines. It 
is contributed by  the fact that the validity of the
approximation does not solely depends on the $EW$ of the OVII line and, to a lesser
extent, by  the presence of inhomogeneities in the absorbing gas.
Although we show only the case of NeIX and OVII, we have checked that
the same trend is found for the other ratios that use the  OVIII and OVI lines.
We note that the $\sim 30$ \% of the absorbers for which the minimization fails
to find a $(\rho_{\rm g},T_{\rm g})$ solution typically have a large $\Delta$. 

The existence of an anti-correlation between $\Delta$ and
$EW_{\rm M}(OVII) $ suggests a practical, although approximate  way to correct  for systematic errors
since one can use the  $\Delta$ - $EW_{\rm M}(OVII) $ relation in reverse
to correct the predicted $EW$ ratio knowing the $EW$ of the OVII line.

To first approximation, this correction should not depend
on the WHIM model since for a given gas density, temperature and relative metal abundance,
the predicted $EW$ ratio depends only on the ionization model. 
Therefore, under the hypothesis of hybrid, photoionization  plus collisional ionization, 
and assuming the same UV and X background
we expect that the mean $\Delta$ - $EW_{\rm M}(OVII) $ relation should
not depend on  the WHIM model.
To check this hypothesis we have repeated the same analysis
using model  $B2$ instead of $B1$
and found that, at least in the range of  $EW_{\rm s}$  we are interested in,
the linear fit to the  mean  $\Delta \ - \ EW_{\rm M}(OVII) $ relation
is consistent, within the errors,  with the one shown in  Fig.~\ref{fig:correction}.

Figs.~\ref{fig:rr} and  ~\ref{fig:tt}
compare the logarithm of the density and temperatures
predicted from the OVII, OVIII and NeIX EW line ratios
to their true values  after applying the statistical  correction
described above. The correction is very effective in reducing
systematic but also random errors.
Systematic errors on the density and the temperature
decrease to $\sim 17$ \% and $\sim 10$ \%, respectively
and are comparable, to  random errors
of $\sim 23$ \% and $\sim 25$ \%.
Similar results were obtained for absorbers featuring the  OVI
rather than the NeIX line. In this case, however, systematic errors
are further reduced  to a few percent level.

To summarize, our analysis shows that one can measure the effective
temperature and density  of an absorption system by
measuring  the EW ratio of at least three lines.
However, the estimate is affected by systematic errors
that are mainly contributed by the breakdown of the optically-thin approximation.
Nonetheless, once  the EW of the OVII line is measured accurately,
unbiased estimates can  be obtained through  simple and
robust corrections that depend on the ionization model adopted.
Applying this technique to  absorption systems
featuring the NeIX line detected in the GRB afterglow spectra
measured by {\small  EDGE/XENIA}, one expects to be able to
measure the density and temperature  of about 16 absorbers per year
with  $\sim 25$ \% random errors. This estimate is obtained by multiplying the
number of detectable absorption systems listed in Table~4 by the
the 70 \% success rate of the estimation technique.

We note that this estimate requires {\it a-priori} knowledge
of  the Ne/O relative abundance.
This needs not to be true when we
focus on those absorbers featuring  the OVI, OVII and OVIII lines. Unfortunately, 
because of the low OVI column density,
the OVI 1s-2p line is very weak and one expects to detect  only $\sim 2$
such absorbers per year, as shown in Table~4. Once more,
to increase the number of OVI detections one should
resort to simultaneous UV spectroscopy.

Finally, we point out that we have only considered Neon and Oxygen ions.
Using all other ions listed in Table~4 would significantly increase the
the number of absorbers with at least three lines for which one can estimate
the temperature and the density of the gas.

\subsection{Measuring the WHIM cosmological density}
\label{sec:cosmo}
Another interesting question is how to estimate the cosmological mass density of the
WHIM from the  line $EW_{\rm s}$ measured in the GRB afterglows  spectra.
As pointed out by Richter \etal (2008) this estimate can be obtained in two steps.
First, following the procedure outlined in the previous Section,
one uses the measured $EW_{\rm s}$  of the ion lines and their ratios to determine
the ion fractions and, from these, the total gas column density for a given metallicity.
Then, one integrates the gas column density along LOS to different
X-ray sources  in order to minimize statistical errors and cosmic variance.
Following this procedure, the cosmological density of the WHIM in units of the critical density $\rho_c$ $\Omega_{\rm WHIM}$,
can be estimated  from all absorption systems featuring the OVII line as:
\beq
\Omega_{\rm WHIM}  = \frac{\mu m_H}{\rho_c}
\frac{
\sum_{i,j} \left(
\frac{A_H}{A_O} \right)_{i,j}
\left(
\frac{N(OVII)}{X(OVII)}
\right)_{i,j}}
{\sum_j d_c(z_j)},
\label{omegawhim}
\eeq
where  $\mu=1.3$ is the mean molecular weight and $m_H$ is the mass of the hydrogen
atom. The index $i$ denotes the absorption systems along the LOS $j$.
Each absorber is characterized by the OVII column density $N_{OVII}$,
that can be evaluated from the line 
$EW$, the ion fraction, $X_{OVII}$, which  can be evaluated from the $EW$ ratios, and
the metallicity $(A_O/A_H)$ that needs to be assumed {\it a priori}.
The quantity $d_c(z_j)$ denotes the  comoving distance corresponding to
the redshift range in which absorption systems can be detected along the line of sight $j$
and depends on the assumed cosmology.
It is straightforward to generalize Eq.~\ref{omegawhim}  to account for all remaining
absorption systems with featuring lines different from 
OVII l which, however,  are very few.

Nicastro \etal (2005) used an estimator similar to that of Eq.~\ref{omegawhim}
to measure $ \Omega_{\rm WHIM} $ from the putative absorption systems along the LOS to Mrk 421.
They assumed that the $(A_O/A_H)$ ratio is constant and that errors are
dominated by Poisson noise from low number statistics.
Under the same hypotheses,  one would expect to estimate $ \Omega_{\rm WHIM} $
with $\sim$ 10 \% accuracy  using the $N_{abs}\sim 50$ absorption systems that
{\small  EDGE/XENIA} is expected to detect in the GRB afterglows spectra during a
3-year observational campaign.

This  error estimate is clearly too optimistic since it does not account for the uncertainties
in the estimated ion fraction and ignores both the cosmic variance and  the fact that
different absorbers may have 
different metallicity.  In an attempt to obtain a more realistic
error estimate we have measured $ \Omega_{\rm WHIM}$ using all  absorption systems
listed in Table~4 through the  following, simple estimator:
\beq
 \Omega_{\rm WHIM}  =
 \frac{ \langle D_{\rm WHIM} \rangle }{\rho_c}
 \frac{\sum_{i,j}  \rho^{g}_{i,j}}
 {\sum_j d_c(z_j)}
 \label{omegasimple}
 \eeq
$\rho^{g}$ represents the effective gas density of the absorber estimated
from the $EW$ ratios using the technique described in the previous Section
and for which we have also estimated the associated errors.
$\langle D_{\rm WHIM} \rangle$ represents the average
comoving depth of the WHIM absorbers. This quantity has been computed using all
 mock absorption systems with a detectable OVII line and represents the
comoving distance corresponding to the redshift interval
$[\underline{z},\overline{z}]$ in which the transmitted OVII flux
is below 95 \%. Clearly, $\langle D_{\rm WHIM} \rangle$ is expected
to depend on the  WHIM model adopted. For model
$B1$ we found that $\langle D_{\rm WHIM} \rangle = 1.4$ Mpc with a
scatter around the mean of  0.6 Mpc that quantifies the random error.
Uncertainties on $\Omega_{\rm WHIM}$ have been obtained 
by adding in quadrature the errors in the measured $\rho^{g}$ 
and the scatter in $D_{\rm WHIM}$ and
and by assuming that the total uncertainty
decreases as $\sqrt N_{abs}$. We neglect the cosmic variance.
Under these hypotheses the errors on $\Omega_{\rm WHIM}$ is $\sim$ 15  \%,
not too different from  the simple Poisson estimate.

Applying Eq.~\ref{omegasimple} to the mock absorption systems
we obtain $\Omega_{\rm WHIM} = 0.014 \pm 0.002$, slightly smaller that the
expected  value $\Omega_{\rm WHIM} = 0.017$. The 1.5 $\sigma$ difference
reflects the fact that  the distribution of $D_{\rm WHIM}$
 is not Gaussian but
is skewed toward large values.

This exercise suggests that  the analysis of
X-ray  GRB spectra taken in 3 years by a satellite like {\small EDGE/XENIA}
should allow to measure $\Omega_{\rm WHIM}$ with 15  \% accuracy.
However, we stress that this estimate strongly depends on the WHIM
models adopted, which is used either to determine the  metallicity of the
absorbers, $(A_O/A_H)$.  in Eq.~\ref{omegawhim} or their average depth,
$\langle D_{\rm WHIM}, \rangle$ in Eq.~\ref{omegasimple}.

\section{Discussion and Conclusions}
\label{sec:discussion}

In this work we have considered different numerical models of the WHIM consistent with currently available observational constraints to assess the possibility
of detecting  the signature of the missing baryons in the X-ray spectra of distant 
sources and to study
their physical properties.
In particular we focused on the absorption spectra of distant GRB afterglows that
could be taken by the Wide Field Spectrometer on board of the recently proposed {\small EDGE} and {\small  XENIA} missions. Both satellite are capable of fast repointing, which make them suitable
for observing the GRB afterglows during their early, bright phase.
We note that although we have only considered absorption spectroscopy
the WFS  is especially suitable for studying the WHIM in emission,
as we will show in a future work.
The main results of this study can be summarized as follows:

{\em (i)} All WHIM  models
explored provide similar predictions for the cumulative distribution
of OVI line $EW$ but not for OVII and OVIII.
This fact reflects the differences in the metallicity prescriptions among models,
which become more evident  in regions of enhanced density and temperature
in which the OVII and OVIII ions  are more common than OVI.
As a consequence the OVI lines are not  efficient signposts
for  the bulk of the WHIM which, instead,  is best traced by OVII and OVIII, as already
shown by  Chen \etal (2003).

{\em (ii)} GRBs are better X-ray background sources than AGN in many aspects.
First of all, for a reference fluence in the [0.3-2] keV band
of  $10^{-6} \ {\rm erg} \ {\rm cm}^{-2}$ required to detect the WHIM,
the typical GRB distance is larger than that of AGN, hence increasing the
probability of WHIM detection. Second of all, GRBs constitute
a replenishable reservoir of sources, contrary to AGN.
As a consequence, the total number of GRBs above the reference fluence
that {\small  EDGE/XENIA} is  expected to detect in a year is already a factor of two
larger that the total number of  available AGN with $z>0.3$ with the further advantage
that the photons can be  collected in  half observational time.

{\em (iii)} A satellite like {\small  EDGE/XENIA} is expected to detect 25-60 WHIM absorbers per year
along the LOS to $\sim 30$ GRBs through the OVII line.
The statistical significance of the detections is 4.5 $\sigma$ which guarantees
that the probability of spurious line contamination is less than 1 \%.
The lower bound corresponds to  the most conservative WHIM model considered here
(dubbed $M$) which, however, we regard as less realistic since it  assumes a deterministic metallicity-density relation.
The upper value derives from a more physically sound WHIM model ($B2$) that accounts for
the scatter in the relation. A more realistic upper bound should also account for the 
possibility of detecting WHIM absorbers beyond $z=1$, which we have ignored 
in our analysis. 
Finally, with a goal energy resolution of 1eV 
instead of 3 eV (baseline) the number of expected detections would further increase
by about 40 \%.

{\em (iv)} A significant fraction of the WHIM absorbers
detected through the OVII line comes with an additional,  detectable OVIII line.
The total number of expected WHIM absorption systems
featuring both the OVII and OVIII line  is  about 35 per year
when we consider  model $B2$.
The simultaneous detection of both lines increases the significance of
the detection to 5 $\sigma$. A large fraction of these systems
have a third, detectable ion line, typically of a different element,
that further increases the significance of the detection.

{\em (v)} Measuring the $EW$ of at least three lines allows  to characterize the thermal
properties of the absorber. Under the condition of hybrid collisional and photoionization equilibrium
adopted in our WHIM models and assuming an optically thin medium,
we were able to determine  the effective density and temperature of the gas
in $\sim $ 70 \% of the line absorption systems featuring a NeIX or OVI line in addition to OVII and OVIII.
These estimates, however, are affected by systematic errors that
derive from the breakdown of the optically thin approximation. A simple, model independent correction
can be applied that allows to remove  this systematic effect.
As a result one should be able
to estimate the logarithm of the temperature and density of about 20 WHIM absorption systems per year with $\sim 25$ \% random errors.

{\em (vi)} Integrating the gas density along the lines of sight to 
the multiple line absorbers that a mission like  {\small  EDGE/XENIA}  is expected to
detect during the first three year of the mission,  would allow to estimate the
cosmological mass density of the WHIM with an accuracy of $\sim$ 15 \%.
This estimate, however, heavily relies on the assumed gas metallicity  which currently
constitutes  the main uncertainties of the WHIM models.

As soon as they will become available, X-ray observations will be used to
validate the WHIM models and therefore  to constrain the poorly known metal 
enrichment mechanisms.
Nevertheless, with no rigorous metallicity model available, the abundance of the heavy ions in
the WHIM can only be measured by observing the Ly-$\alpha$ hydrogen line
associated to the X-ray absorbers.
The {\it Cosmic Origin Spectrograph} will allow to search for
broad Ly-$\alpha$ absorbers  which are likely associated to the WHIM
along the LOS to bright AGN. For transient sources, like
GRBs, a similar study would require the possibility of performing
simultaneous X-ray and UV spectroscopy which is currently not envisaged in the 
{\small EDGE} or {\small  XENIA} mission concepts.
Yet,  the success of a combined
UV and X-ray spectroscopy strategy strongly depends
on how well broad Ly-$\alpha$  absorbers trace the WHIM gas.
Current datasets provide no clue on their mutual relation.
On the theoretical side  Chen \etal (2003)
found weak correlation between HI and  both OVII and  OVIII that, as 
we have stressed already,  trace the bulk of the WHIM.
However, a dedicated  theoretical study must be performed 
to assess how well  a combined UV and X-ray spectroscopy will
measure the metal content of the WHIM gas.

\begin {figure}
\centering
\epsfig{file=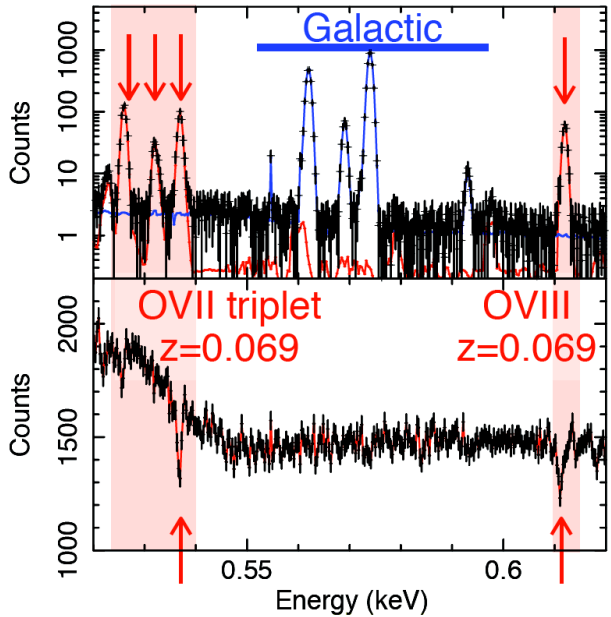,width=0.45\textwidth}
\caption{Bottom panel. Zoomed plot in a narrower energy range
of the mock absorption spectrum in Fig.~\ref{fig:convolved}, centered on the OVII - OVIII absorption
systems at $z=0.069$. Top panel: Emission spectrum  in a 4 armin$^2$ area along the same direction,
taken in a 1 Ms observation with {\small EDGE/XENIA} Wide Field Spectrograph. The OVII triplet
and the OVIII resonant line associated to the absorption lines are  clearly seen along with the
Galactic OVII triplet.}
\label{fig:amabs}
\end{figure}

Although we did not mention it explicitly,
the Wide Field Spectrometer on board of  {\small EDGE} or {\small  XENIA}
offers the possibility of combining emission and absorption analyses.
This is an important asset of the {\small EDGE/XENIA} mission concepts since detecting
the OVII resonant line in absorption and the triplet in emission 
allows to constrain the metal abundance of the WHIM.
As in the case of the combined UV and X-ray spectroscopy, 
the success of this alternative strategy
depends on how well WHIM absorbers can trace the emitting gas
and on our ability to remove or model the contribution of the different backgrounds/foregrounds.
The combined study of the WHIM in absorption and emission
is the subject of  future work. Here we anticipate that
on average ~30 \% of the line systems detected in absorption can
also be seen in emission in deep observations.
Figure~\ref{fig:amabs} illustrates one of these cases. The lower panel shows a zoom in
a narrow energy interval of the mock spectrum displayed in Fig.~\ref{fig:convolved}, centered
on the OVII and OVIII lines of the absorption system at $z=0.069$.
The upper panel shows the emission spectrum in a 4 armin$^2$ area taken
along the same LOS when the GRB is off in a 1 Ms long observation.
The OVIII and OVII lines are detected both in absorption and in emission.
All components of the OVII triplet are clearly seen and can be
well separated from the Galactic component.
The combined emission/absorption spectroscopy not only  constitutes
a possible way to constrain the WHIM metallicity and
hence to estimate its cosmological density, but also suggests an observational strategy
for the WHIM which consists in  pre-selecting target fields in which the
WHIM absorption lines have been detected in the X-ray spectrum of a GRB for deeper, follow-up observations once the GRB has faded away.

The possibility of studying the WHIM in emission is crucial to investigate its spatial distribution, since
the pencil-beam-like  sampling provided by absorption spectroscopy is too sparse to
trace its large scale, filamentary structure that we observe in the simulations.
Yet, cross-correlation of WHIM absorbers
and next generation all-sky galaxy surveys  out to $z=2$ like {\small ADEPT} or {\small EUCLID}
may shed light on the WHIM-galaxy connection, very much like in the recent correlation analyses
of weak Ly-$\alpha$ absorbers and HI-rich galaxies performed by Ryan-Weber (2006) and Pierleoni \etal (2008) and that of OVI absorbers and galaxies (Ganguly \etal 2008). These studies 
have helped in clarifying
the relation between the diffuse baryons in the intergalactic medium and the 
atoms locked inside virialized structures.

Finally, we would like to stress that although X-ray spectroscopy constitutes the best
way to study the missing baryons, alternative strategies like the cross correlation
of the kinetic Sunayev-Zel'dovich signal near clusters and groups
and the all-sky E-mode polarization in the context of upcoming CMB experiments
like {\small Planck}, {\small ACT}, {\small SPT} or {\small APEX}
will provide useful, independent constraints on the presence and abundance of
missing baryons in the Universe   (Hernandez-Monteagudo \& Sunyaev 2008).

\section{Acknowledgments}

We are grateful to Stefano Borgani for providing us  the outputs of 
Borgani \etal (2004) hydrodynamical simulation, his many comments and helpful 
suggestions about how we could improve the manuscript.
We also thank F. Fiore,  P. Mazzotta and C. Perola for useful discussions.
We acknowledge financial contribution from contract ASI-INAF I/088/06/0.
SRON is supported financially by NWO, the Netherlands Organization for Scientific Research.

\end{document}